\documentclass[smallextended]{svjour3}       % onecolumn (second format)
\usepackage[numbers]{natbib}
\usepackage{url}

\smartqed  % flush right qed marks, e.g. at end of proof
\usepackage{graphicx}
\usepackage{comment}
\usepackage{amssymb}
\usepackage{ifthen}
\usepackage{paralist, tabularx}
\usepackage{listings}
\usepackage{tabularx}
\usepackage{booktabs}
\usepackage{siunitx}
\usepackage{caption}
\usepackage{multirow}
\usepackage{tabularx}
\usepackage{comment}

\usepackage{tcolorbox}

\newcommand{\COMMENT}[1]{}

\usepackage{xspace}

\newcommand{\etc}{etc.\xspace}
\newcommand{\etal}{\emph{et~al.}\xspace}

\newcommand{\listref}[1]{Listing~\ref{#1}\xspace}

\newtcolorbox{resultbox}{colback=gray, arc=0.5mm, top=1mm, bottom=1mm, left=1mm, right=1mm}

\newboolean{showcomments}

\definecolor{codegreen}{rgb}{0,0.6,0}
\definecolor{codegray}{rgb}{0.5,0.5,0.5}
\definecolor{codepurple}{rgb}{0.58,0,0.82}
\definecolor{backcolour}{rgb}{0.95,0.95,0.92}
\definecolor{javapurple}{rgb}{1,0.5,0.31} % keywords
\definecolor{javablue}{rgb}{0.25,0.35,0.75}
\definecolor{backbefore}{rgb}{0.94,0.63,0.63}
\definecolor{backafter}{rgb}{0.6,1,0.6}

\definecolor{gray}{rgb}{0.83,0.83,0.83}
\definecolor{green}{rgb}{0.56,0.93,0.56}

\setboolean{showcomments}{true}

\ifthenelse{\boolean{showcomments}}
{\newcommand{\nb}[2]{
		\fbox{\bfseries\sffamily\scriptsize#1}
		{\sf\small$\blacktriangleright$\textit{#2}$\blacktriangleleft$}
	}
}
{\newcommand{\nb}[2]{}
}

\newcommand\LEUSON[1]{\textcolor{magenta}{\nb{LEUSON}{#1}}}

\newcommand\PHILIPPE[1]{\textcolor{blue}{\nb{PHILIPPE}{#1}}}

\newcommand{\listComp}{\textit{List Comprehension}}
\newcommand{\setComp}{\textit{Set Comprehension}}
\newcommand{\dictComp}{\textit{Dictionnary Comprehension}}
\newcommand{\chainComparison}{\textit{Chain Compare}}
\newcommand{\truthValueTest}{\textit{Truth Value Test}}
\newcommand{\loopElse}{\textit{For-Else}}
\newcommand{\assignMultiTargets}{\textit{Assign-Multiple-Targets}}
\newcommand{\starInFunctionCall}{\textit{Star-in-Func-Call}}
\newcommand{\forMultiTarget}{\textit{For-Multiple-Targets}}

\usepackage{listings}

\lstdefinestyle{mystyle}{backgroundcolor=\color{backcolour},commentstyle=\color{codegreen},
  keywordstyle=\color{javapurple},
  numberstyle=\tiny\color{codegray},
  stringstyle=\color{codepurple},
  basicstyle= \textnormal,
  breakatwhitespace=false,         
  breaklines=true,                 
  captionpos=b,                    
  keepspaces=true,                 
  numbers=left,                    
  numbersep=5pt,                  
  showspaces=false,                
  showstringspaces=false,
  showtabs=false,                  
  tabsize=2,
}
\newcommand{\rqone}{What is the prevalence of performance smells in ML systems vs non-ML systems ?}
\newcommand{\rqtwo}{What are the different types of performance smells in ML systems ?}
\newcommand{\rqthree}{How are performance smells distributed across the stages of the ML pipeline?}

\begin{document}

\title{Performance Smells in ML and Non-ML Python Projects: A Comparative Study}


\author{François Belias         \and
        Leuson Da Silva \and Foutse Khomh \and Cyrine Zid
}


\institute{François Belias  \and Leuson Da Silva \and Foutse Khomh \and Cyrine Zid \\
        Department of Computer Engineering and Software Engineering \\
        Polytechnique Montreal \\
        Montreal, QC, Canada \\
        \email{\{francois-philippe.ossim-belias, leuson-mario-pedro-2.da-silva, foutse.khomh, cyrine-2.zid\}@polymtl.ca}}

 \date{}

\maketitle

\begin{abstract}
Python is widely adopted across various domains, especially in Machine Learning (ML) and traditional software projects. Despite its versatility, Python is susceptible to performance smells, i.e., suboptimal coding practices that can reduce application efficiency. 
This study provides a comparative analysis of performance smells between ML and non-ML projects, aiming to assess the occurrence of these inefficiencies while exploring their distribution across stages in the ML pipeline. 
For that, we conducted an empirical study analyzing 300 Python-based GitHub projects, distributed across ML and non-ML projects, categorizing performance smells based on the RIdiom tool. 
Our results indicate that ML projects are more susceptible to performance smells %, particularly for \textit{Assign Multi Targets}, \textit{Truth Value Test}, and \textit{List Comprehension}, 
likely due to the computational and data-intensive nature of ML workflows. 
We also observed that performance smells in the ML pipeline predominantly affect the Data Processing stage. 
However, their presence in the Model Deployment stage indicates that such smells are not limited to the early stages of the pipeline.
%We also observed that these performance smells on the ML pipeline affect the \textit{Data processing} stage the most. In contrast, the \textit{Model Deployment} stage shows that performance smells are not restricted to the early stages of the ML pipeline.
Our findings offer actionable insights for developers, emphasizing the importance of targeted optimizations for smells prevalent in ML projects. 
Furthermore, our study underscores the need to tailor performance optimization strategies to the unique characteristics of ML projects,  with particular attention to the pipeline stages most affected by performance smells. % and non-ML Python projects.
\end{abstract}
\keywords{Python \and Code Smells \and Performance Smells \and Python Idioms \and Empirical study \and Machine learning \and Traditional systems}

\section{Introduction} \label{sec:intro}
Python’s versatility and rich ecosystem have cemented its popularity in software development, particularly for Machine Learning (ML) systems, where it supports complex tasks like Model Training and Data Processing \cite{peng2021empirical, dyer2022exploratory}. However, despite its benefits, Python can pose challenges, especially around performance efficiency and security \cite{alfadel2023empirical, paramitha2023technical}. Performance inefficiencies, or \textit{performance smells}, refer to coding practices that reduce execution efficiency, affecting critical metrics such as response time, memory usage, and throughput \cite{4907670}. %\LEUSON{The next sentence should be placed in the paragraph when you discuss the impact of ML in energy consume. But before discussing this topic, you have to discuss the impact of non-ML python projects in energy consume. You might also bring some references about this topcic (like, https://dl.acm.org/doi/pdf/10.1145/3139367.3139418). Then, you can say that such a problem is even worst when it comes to ML, and discuss, like you did next.} While performance issues are a general concern, they are especially problematic in ML projects, which often handle large datasets and high computational loads \cite{wan2019does}.

Beyond software quality concerns, performance inefficiencies in Python-based systems also raise significant energy consumption issues. Even in non-ML projects, inefficient code can lead to unnecessary hardware usage and increased energy demands. For instance, Pereira et al. \cite{georgiou2017analyzing} show that Python projects can exhibit substantial energy consumption patterns depending on coding practices and language features used.

%Beyond software quality concerns, performance inefficiencies in Python-based machine learning (ML) systems have broader implications in terms of energy consumption and environmental sustainability. The rapid growth of ML has been accompanied by an increasing demand for computational resources, particularly for tasks such as large-scale data preprocessing and model training, which are often executed on high-performance hardware like GPUs and TPUs \cite{7723730}. These operations, while essential, can consume significant amounts of electricity, contributing to the growing carbon footprint of the software industry \cite{strubell2019energypolicyconsiderationsdeep}. For example, training state-of-the-art models like BERT has been shown to result in CO$_2$ emissions comparable to transatlantic flights \cite{strubell2019energypolicyconsiderationsdeep}, highlighting the pressing need to make ML development more resource-conscious.

In this context, performance smells are not only detrimental to execution speed or memory usage but also represent hidden contributors to energy waste. Although a single inefficient loop or redundant comparison might appear harmless in isolation, ML workflows often amplify these inefficiencies due to the sheer volume of iterations and data processed. As a result, such smells can lead to prolonged training durations and excessive hardware utilization. Prior studies show that inefficiencies in Python code can directly influence energy profiles even outside the ML context \cite{georgiou2017analyzing}, which underlines the importance of identifying and mitigating them wherever possible. Identifying and refactoring such smells, particularly in computation-intensive stages like Data Processing, is therefore critical not just for performance optimization, but also for supporting the broader goals of green software engineering \cite{kuchnik2022plumber}. Green software engineering promotes environmentally sustainable software development by reducing the energy and hardware resources consumed throughout the software lifecycle. Improving performance efficiency is one of its core strategies \cite{abraham2003green}. %\LEUSON{(addressed: expanded on green software engineering)}

%In this context, performance smells are not only detrimental to execution speed or memory usage but also represent hidden contributors to energy waste. Inefficient loops or inefficient comparisons can lead to unnecessary hardware occupation, longer training times, and ultimately greater environmental impact. \LEUSON{I think here it's important to highligh that a single loop won't be responsible for much energy consumption, but the context around ML implies such a high consmuption. You can even rely again on references that show that Python projects consume energy etc. }Identifying and refactoring such smells, particularly in computation-intensive stages like Data Processing, is therefore critical not just for performance optimization, but also for supporting the broader goals of green software engineering \cite{kuchnik2022plumber}. %\LEUSON{Add reference.}

%\LEUSON{Since we have space now, I think we can discuss the green software engineering concept. For example, improving performance is a way to achieve green SE goals. You can start with the overall idea, and then discuss the impact observed by ML projects, considering their specificities; for example, dealing with more data, processing, etc. So, you would move the last sentence of the previous paragraph here and expand.}

Despite the high stakes, little is known about the relationship between performance smells and the specific challenges of ML development. Existing research on Python tends to emphasize general code quality, with limited focus on how performance smells specifically impact execution speed and memory in both ML and non-ML contexts \cite{gesi2022codesmellsmachinelearning}. While prior studies have focused on code maintainability and structural design issues \cite{Khomh12, 10.1007/3-540-45672-4_31, palomba2018diffuseness}, few have systematically investigated how inefficiencies manifest in ML pipelines, or how they differ from those in general-purpose Python applications. Given the high computational demands of ML systems, overlooking such inefficiencies may lead to significant energy and performance losses.

Our study seeks to fill this gap by systematically analyzing the occurrence of performance smells in ML versus non-ML projects. By identifying common performance smells and their frequency across diverse Python applications, we aim to provide developers with actionable insights to improve performance, reduce resource consumption, and optimize Python’s use in varied project types. Overall, in this study, we aim to address a critical need for understanding and managing performance bottlenecks, ultimately supporting more efficient Python-based ML pipelines and other applications. Specifically, we answer the following research questions:
\begin{itemize}
    \item{\bf RQ$_1$:} \rqone
    \item{\bf RQ$_2$:} \rqtwo
    \item{\bf RQ$_3$:} \rqthree
\end{itemize}

For RQ$_1$, we aim to quantify the presence of performance smells across different types of Python repositories. Understanding whether ML projects are more prone to such issues can help prioritize performance-oriented tooling and educational efforts.
With RQ$_2$, by identifying the most frequently occurring performance smells, we can highlight specific coding patterns that are especially problematic in ML development. 
This may guide optimization efforts and inform best practices tailored to ML codebases.
Finally with RQ$_3$, we explore the distribution of performance smells across ML pipeline stages. 
It helps pinpoint where developers should concentrate their optimization efforts to achieve the most significant performance and energy efficiency gains. %\LEUSON{Well done.}

Our results highlight the prevalence of performance smells in ML projects compared to non-ML projects. 
Specifically, ML projects demonstrate a higher mean frequency of performance smells, particularly for certain types such as \textit{Assign Multi Targets}, \textit{Truth Value Test}, and \textit{Chain Compare}. 
Such a difference might be attributed to the nature of ML projects, which often involve manipulating large datasets and high computation tasks, leading to increased occurrences of these antipatterns. 
Moreover, our findings reveal that the Data processing stage is the most affected by the performance smells analyzed. In contrast, the Model Deployment stage shows that performance
smells are not restricted to the early stages of the ML pipeline.

These findings carry several implications. From a developer’s perspective, understanding the specific smells and their typical locations within the ML workflow can lead to more targeted performance optimizations. For researchers and tool builders, our results suggest the need for stage-aware linters and static analysis tools that focus on ML-specific coding practices. From a sustainability standpoint, reducing such inefficiencies can lead to lower resource usage and decreased energy consumption, aligning with the principles of green software engineering \cite{ANASTAS200611}.

%\LEUSON{Discuss a bit about the implications.}

The remainder of the paper is structured as follows. Section \ref{sec:background} describes the performance smells under study. Section \ref{sec:design} details our dataset collection approach. Sections \ref{sec:RQ1}, \ref{sec:RQ2}, \ref{sec:RQ3} present the methodology and findings for each research question. Section \ref{sec:research_guidelines} present the guidelines of our study, Section \ref{sec:threats} discusses threats to validity, Section \ref{sec:related} reviews related work, and Section \ref{sec:conclusion} summarizes our findings and outlines future directions.

\section{Background}
\label{sec:background}
%\subsection{Performance smells}
%\label{sub_sec:Performance_smells}
%\lm{If we keep the description/explanation of each performance smell analyzed in the study, we might move it to a Background section (after Introduction).}
This study focuses on a catalogue of nine performance smells that arise from non-idiomatic Python practices. 
These smells, identified by Zhang \etal \cite{zhang2023faster}, represent patterns that lead to performance issues in Python-based systems. 
Understanding these smells is essential for improving code performance and preventing resource-intensive bottlenecks, particularly in computationally demanding environments like ML projects.
Next, we provide a detailed overview of each performance smell.
%\LEUSON{I wonder whether you could provide the code before and after like you do for some cases.}
\begin{enumerate}

\item \listComp
    \begin{lstlisting}[language=Python, caption= Using a for loop to add element in a list ,label=lst:anti_idiom_list_Comp,basicstyle=\small\ttfamily] 
    a = [] 
    for e in range(10): 
        a.append(e) \end{lstlisting}
    
    \begin{lstlisting}[language=Python, caption= Filling a list using list comprehension syntax, label=lst:list_comp, basicstyle=\small\ttfamily] 
    a = [e for e in range(10)] \end{lstlisting}

In \listref{lst:anti_idiom_list_Comp}, the list is built using a \emph{for} loop and \emph{append()}, which is more verbose and less efficient than using a \emph{list comprehension}, as shown in \listref{lst:list_comp}.

\item \setComp
    \begin{lstlisting}[language=Python, caption= Using a for loop to add element in a set ,label=lst:anti_idiom_set_Comp,basicstyle=\small\ttfamily]
    b = set() 
    for e in range(10): 
        b.add(e) \end{lstlisting}
    
    \begin{lstlisting}[language=Python, caption= Filling a set using set comprehension syntax, label=lst:set_comp, basicstyle=\small\ttfamily] 
    b = {e for e in range(10)} \end{lstlisting}

In \listref{lst:anti_idiom_set_Comp}, the set is built using a \emph{for} loop and \emph{.add()}, which is suboptimal compared to Python's \emph{set comprehension} capabilities, as shown in \listref{lst:set_comp}.

\item \dictComp
    \begin{lstlisting}[language=Python, caption= Using a for loop to add element in a dictionary ,label=lst:anti_idiom_dict_Comp,basicstyle=\small\ttfamily] 
    b = {} 
    for k, v in a.items(): 
        b[k] = v \end{lstlisting}
    
    \begin{lstlisting}[language=Python, caption= Filling a dictionary using dictionary comprehension syntax ,label=lst:dict_comp,basicstyle=\small\ttfamily] 
    b = {v: k for k, v in a.items()} \end{lstlisting}
    
    \listref{lst:anti_idiom_dict_Comp} builds a dictionary by manually assigning key-value pairs. Python’s \textit{dictionary comprehension} capabilities provide a cleaner and often faster alternative, as seen in \listref{lst:dict_comp}.

\item \chainComparison
    \begin{lstlisting}[language=Python, caption= Using separate comparisons in a chain ,label=lst:anti_idiom_chain_Comp,basicstyle=\small\ttfamily]
    i > n1 and i <= n1 + n2 \end{lstlisting}
    
    \begin{lstlisting}[language=Python, caption= Efficient chain comparison,label=lst:idiom_chain_Comp,basicstyle=\small\ttfamily] 
    n1 < i <= n1 + n2 \end{lstlisting}
    
    In \listref{lst:anti_idiom_chain_Comp}, separate comparisons are used. This can be made more readable and idiomatic using Python's built-in chained comparison syntax, as shown in \listref{lst:idiom_chain_Comp}.

\item \truthValueTest
    \begin{lstlisting}[language=Python, caption= Explicit checking truth values ,label=lst:anti_idiom_tvt,basicstyle=\small\ttfamily] 
    if n % 2 != 0: 
        pass \end{lstlisting}
    
    \begin{lstlisting}[language=Python, caption= Efficient checking truth values ,label=lst:idiom_tvt,basicstyle=\small\ttfamily] 
    if n % 2: 
        pass 
    \end{lstlisting}

In \listref{lst:anti_idiom_tvt}, the expression explicitly checks for a non-zero value. Python allows direct truth value testing, making the condition more concise, as shown in \listref{lst:idiom_tvt}.

\item \loopElse
    \begin{lstlisting}[language=Python, caption= Using additional flag to check the for-loop exit ,label=lst:anti_idiom_LE,basicstyle=\small\ttfamily] 
    finishedForLoop = True 
    for x in range(2, n): 
        if n % x == 0: 
            finishedForLoop = False 
            break 
    if finishedForLoop: 
        pass \end{lstlisting}
    
    \begin{lstlisting}[language=Python, caption= Track loop has finished using for For-Else ,label=lst:for_else,basicstyle=\small\ttfamily] 
    for x in range(2, n): 
        if n % x == 0: 
            break 
    else: 
        pass \end{lstlisting}

In \listref{lst:anti_idiom_LE}, an additional flag is used to track loop termination. This logic can be simplified using Python's \emph{for-else} construct, as shown in \listref{lst:for_else}.

\item \assignMultiTargets
\begin{lstlisting}[language=Python, caption=Using multiple assignments statements ,label=lst:anti_idiom_AMT,basicstyle=\small\ttfamily] 
f = d[0]
d[0] = d[e]
d[e] = f \end{lstlisting}

\begin{lstlisting}[language=Python, caption=Efficient way to assign values ,label=lst:AMT,basicstyle=\small\ttfamily] 
d[0], d[e] = d[e], d[0] \end{lstlisting}

In \listref{lst:anti_idiom_AMT}, each attribute is assigned to its corresponding value individually. While this approach is functional, it is verbose and less efficient and readable than Python's feature to perform \emph{multiple assignments} in a single statement as in \listref{lst:AMT}.

\item \starInFunctionCall
    \begin{lstlisting}[language=Python, caption=function call with explicit multiple arguments,label=lst:anti_idiom_SIFC,basicstyle=\small\ttfamily] 
    dicts = load_crowdhuman_json(sys.argv[1], sys.argv[2], sys.argv[3]) \end{lstlisting}
    
    \begin{lstlisting}[language=Python, caption=Passing arguments using star operator,label=lst:call_star,basicstyle=\small\ttfamily] 
    dicts = load_crowdhuman_json(*sys.argv[1:4]) \end{lstlisting}

\listref{lst:anti_idiom_SIFC} explicitly lists multiple arguments. The code can be made more concise and flexible using the \emph{star operator} to unpack arguments, as in \listref{lst:call_star}.

\item \forMultiTarget
    \begin{lstlisting}[language=Python, caption=iterates through a collection and manually extracts specific elements listed,label=lst:anti_idiom_FMT,basicstyle=\small\ttfamily] 
    for item in sales: a = item[0], item[1] \end{lstlisting}
    
    \begin{lstlisting}[language=Python, caption=Efficient way to iterates through a collection and manually extracts specific elements listed,label=lst:FMT,basicstyle=\small\ttfamily] 
    for e_0, e_1, *e in sales: 
        a = e_0, e_1 \end{lstlisting}

In \listref{lst:anti_idiom_FMT}, elements are manually accessed via indexing. Unpacking directly in the loop header, as shown in \listref{lst:FMT}, improves clarity and efficiency, making the code more Pythonic.

\end{enumerate}
Each anti-idiom reflects practices that lead to less efficient code and should be avoided for optimized Python development. While functional, these patterns do not take full advantage of Python's idiomatic constructs, commonly known for simplifying code and improving both performance and readability.

\section{Dataset Collection}
\label{sec:design}
This study provides insight into the distribution of performance smells between ML and non-ML projects. 
To construct the sample of projects to be analyzed, we adopted a method similar to that of Hadhemi \etal \cite{Hadhemi}, using the GitHub API to select ML and non-ML projects.
In the following sections, we outline the design of our empirical study, detailing the steps taken to select and filter our sample.

\subsection{Machine Learning Projects}
Since we aim to establish a representative sample of ML projects and given the diversity of application domains ML is applied to, we first define a set of relevant domains.
Then, we proceed with selecting and filtering projects accordingly.

\subsubsection{Choice of ML domains} 
Given the diversity within ML projects~\cite{dargan2020survey, deng2016deep, 10.5555/3360273, singh2021applications}, we aimed to ensure a relevant and representative sample of projects for our study.
To this end, we adopt the same domain-based sampling strategy as Bhatia \etal~\cite{bhatia2023empirical}, by defining a set of ML application domains and selecting projects using associated domain-specific keywords.
These domains include {\bf{Image Processing}}, {\bf{Natural Language Processing (NLP)}}, {\bf{Audio Processing}}, {\bf{Autonomous Gameplay}}, and {\bf{Self Driving Car}}.
We also include projects that do not fall into the previous categories but are still relevant, grouped under the {\bf{Other ML}} category, which follows the same categorization as in Bhatia \etal~\cite{bhatia2023empirical}. These typically correspond to educational and assignment-based ML projects.
Including these projects is important, as they represent the initial exposure of developers to ML development and foundational topics, providing information on early-stage coding practices~\cite{bhatia2023empirical, hameed2019review}.

\begin{comment}
    We also include projects that don't fit the previous categories but remain relevant, classified under the Other ML category, following the categorization established by Bhatia et al. \cite{bhatia2023empirical}. This category primarily encompasses educational ML projects, programming assignments, and specialized ML applications such as ML-based computer vision.
The inclusion of educational projects and assignments is crucial, as they represent developers' first exposure to ML development and fundamental concepts, offering valuable insights into coding practices adopted during early learning stages \cite{bhatia2023empirical, hameed2019review}.
\end{comment}

By including ML projects from both open-source and educational contexts, we ensure a broad coverage of development practices and code characteristics.
Combined with the diversity of application domains (e.g., NLP, vision, audio), this variety strengthens the generalizability of our findings on performance smells.
As a result, we provide insights that can help developers identify common performance pitfalls and adopt strategies to enhance code efficiency across a wide variety of ML development scenarios.

\subsubsection{Gathering Repositories}
Once we have defined the target ML domains for our analysis, we started collecting ML projects using the GitHub API.
To narrow the search to relevant projects, we adopted the same querying strategy as Bhatia \etal~\cite{bhatia2023empirical}.
Specifically, for each domain (as shown in Table~\ref{tab:ml_domains}), we used their associated keywords combined with the term \texttt{machine learning} to query GitHub. For each query, we retained the top 50 repositories, as done in their study.

When using the GitHub API search, results are sorted by best match in descending order, unless another sort option is provided as a query parameter.
Multiple factors, such as project description, associated topics and README content, are combined to bring the most relevant items to the top of the result list\footnote{\url{https://docs.github.com/en/rest/search/search?apiVersion=2022-11-28}}. Although stars and forks are not the primary factors in GitHub's ranking algorithm, projects with higher stars and forks tend to rank higher, as these metrics are indicators of popularity and community interest. 
In the end, after the entire process, we collected 1221 projects across all ML domains. Table \ref{tab:ml_domains_dist} presents in detail the distribution of the projects collected by domains. It is important to note that our goal is to retain the top 50 repositories per query, as mentioned before. However, this is not always possible due to our selection criteria, particularly the requirement that projects must be recent. In the case of the \textbf{self Driving Car} domain, we observe that the number of available projects is fewer than 50, which is why we cannot meet the target in this instance.

\begin{comment}
\begin{table}[h]
    \centering
    \begin{tabular}{|l|r|}
        \hline
        \textbf{Domain} & \textbf{Count} \\ \hline
        Self-driving car & 45 \\ \hline
        Audio processing & 376 \\ \hline
        Autonomous gameplay & 200 \\ \hline
        Image processing & 300 \\ \hline
        Natural language processing & 200 \\ \hline
        Other ML & 100 \\ \hline
        Total ML & 1221 \\ \hline
    \end{tabular}
    \caption{ML Domains and their counts}
    \label{tab:ml_domains_dist}
\end{table}
\end{comment}

% Requires: \usepackage{multirow}
\begin{table}[h]
    \centering
    \begin{tabular}{|c|l|p{10cm}|}
        \hline
        \textbf{Projects} & \textbf{Domain} & \textbf{Keywords} \\ \hline
        \multirow{6}{*}{ML} 
        & Self-driving Car & autonomous driving machine learning, self-driving machine learning \\ \cline{2-3} 
        & Audio Processing & sound machine learning, audio machine learning, voice machine learning, speech machine learning \\ \cline{2-3} 
        & Autonomous Gameplay & game reinforcement learning, game machine learning \\ \cline{2-3} 
        & Image Processing & image processing machine learning, vision machine learning, image processing deep learning \\ \cline{2-3} 
        & Natural Language Processing & natural language processing machine learning, chatbot machine learning \\ \cline{2-3}
        & Other ML & machine learning \\ \hline
        \multicolumn{1}{|l|}{Non-ML} & Non-ML & server, database, networking \\ \hline
    \end{tabular}
    \caption{Overview of Projects and Domains}
    \label{tab:ml_domains}
\end{table}

\begin{table}[h]
    \centering
    \begin{tabular}{lcc}
        \hline
        Domains & Before filtering & After filtering \\
        \hline
        Self Driving Car & 45 & 1 \\
        Audio Processing & 376 & 36 \\
        Autonomous Gameplay & 200 & 22 \\
        Image Processing & 300 & 35 \\
        Natural Language Processing & 200 & 11 \\
        Other ML & 100 & 53 \\
        Total ML & 1221 & 158 \\
        \hline
    \end{tabular}
    \caption{Distribution of ML projects before and after automated filtering}
    \label{tab:ml_domains_dist}
\end{table}

\subsubsection{Filtering Projects}
Once we have collected the candidate projects for our study, we further analyze them to filter out those irrelevant. 
For that, we applied a two-level filtering process consisting of automated and manual checks.

\textbf{\emph{Automated Filtering:}}
Previous studies have adopted different approaches to filtering their samples, such as checking the number of stars to estimate the project maturity \cite{dabic2021samplingprojectsgithubmsr}, the number of commits \cite{10.1007/s10664-024-10523-y}, or a combination of these different metrics.
Based on them, we defined a set of criteria to filter our sample as reported in Table \ref{tab:filtering_criteria}; for each criterion, we also present a detailed description. 
As a result of this step, 1062 projects were filtered out, leaving 158 remaining as we can in Table \ref{tab:ml_domains_dist}.
\begin{comment}
\begin{table}[h]
    \centering
    \begin{tabular}{|l|r|}
        \hline
        \textbf{Domain} & \textbf{Count} \\ \hline
        Self-driving car & 1 \\ \hline
        Autonomous gameplay & 36 \\ \hline
        Audio processing & 22 \\ \hline
        Image processing & 35 \\ \hline
        Natural language processing & 11 \\ \hline
        Other ML & 53 \\ \hline
        Total ML & 158 \\ \hline
    \end{tabular}
    \caption{ML Domains and their counts after automated filtering}
    \label{tab:ml_domains_filtered}
\end{table}
\end{comment}
%\LEUSON{You can merge tables 2 and 3.}

\begin{table}[htbp]
\caption{Filtering Criteria for project sampling}
\centering
\begin{tabularx}{\textwidth}{|l|X|X|}
\hline
\textbf{Criterion} & \textbf{Description} & \textbf{Rationale} \\ \hline
\textbf{C1: No Forks} & A project must not be a fork of another project. & Such a criterion avoids duplication and ensures the originality of the codebase, as forks often replicate their main parent project's structure \cite{zampettiempirical}. \\ \hline
\textbf{C2: At least 1 star and 1 Fork} & A project must have at least 1 star and 1 fork. & Such a constraint allows us to remove toy projects while selecting mature ones.
\\ \hline
\textbf{C3: Minimum 5 Source Files} & A project should have at least five source code files. & Projects with fewer files may be immature or underdeveloped. A minimum of five files indicates a certain level of complexity and maturity \cite{bhatia2023empirical}.
\\ \hline
\textbf{C4: 1+ Month Development History} & A project should have more than one month of development history. & This criterion helps to exclude short-term experimental projects that may not offer substantial insights. A month of development typically signals more meaningful contributions to the codebase \cite{tidjon2022empirical}. \\ \hline
\textbf{C5: Recent Activity} & A project should have at least one commit from January 1, 2023, or later (data was collected in August 2024).
& We ensure the project is still active and under development. \\ \hline
\textbf{C6: ML Library Imports} & A project must include at least one file that imports machine learning libraries (e.g., TensorFlow, Keras, PyTorch). & Such a criterion ensures that the project is focused on machine learning, aligning it with our study's goal. \\ \hline
\end{tabularx}
\label{tab:filtering_criteria}
\end{table}

\textbf{\emph{Manual Filtering:}} Despite applying the aforementioned automated filtering criteria to retrieve ML-specific projects from GitHub, we may still encounter projects that are not strictly focused on ML. To ensure that the selected projects were valid for our study, we performed a manual analysis, which involved three researchers working in pairs.
The manual verification process was as follows:
\begin{itemize}
    \item For each project, two researchers independently examined its associated README file. They evaluated the project's purpose and applicability to ensure it was related to machine learning.
    \item If the README did not provide conclusive information, the researchers examined filenames in the repository to check whether they corresponded to common stages of the ML pipeline. For example, files such as \texttt{train.py} and \texttt{data\_collection.py} suggest involvement in training and data collection, respectively.
    \item If the previous steps were not conclusive, the project was excluded from the dataset and considered unrelated to ML.

    \begin{comment}    
    \LEUSON{If not, what happens? The project is classified as non-ML? }
    \PHILIPPE{ I am just going to remove this part as I did not do the computation for that. The idea was to be broad as possible to justify the generability of our study not to do a study per domain/category as the reviewer thought. I will also change the text for any reference implying that}
    \item \textbf{Project Domain:}
    For the projects aligned with the ML domain, the reviewers assessed their objectives and classified them into distinct categories such as personal project, tool, library, course, frameworks, \etc \cite{bhatia2023empirical}. \LEUSON{Motivation for this? Are we going to present the occurrence of smells and comparison across the different domains listed here? I think now we have the space to present this information and enrich the discussion.} 
    \end{comment}
\end{itemize}

Initially, two researchers evaluated individually a list of projects. 
In case of conflicting evaluation, a third reviewer was asked to join the classification, performing a new analysis. 
If the conflict persists, the reviewers discuss it together to reach an agreement.

To evaluate the agreement in our classification, we calculated Cohen's Kappa score \cite{cohen1960coefficient}. The resulting coefficient of 0.98 reflects an almost perfect level of agreement.

After completing this process, we compiled a final list of \textbf{150 projects}, which we selected as the sample of ML projects for our study. Table \ref{tab:domain_projects} describes the selected projects along with their corresponding total number
 of Python source code files per domain.
%\LEUSON{Please, do not place the tables in the middle of a paragraph.}
As shown in Table~\ref{tab:domain_projects}, the distribution of projects across ML domains is uneven. This imbalance is primarily due to our selection criteria, which focused on collecting \textit{recent and active} projects. These constraints naturally filtered out several older or less frequently updated domains such as \textbf{Self Driving Car} and \textbf{NLP}. In contrast, the \textbf{Other ML} category includes many educational and course-related repositories that are both recent and widely shared, especially in the context of online learning platforms or university coursework. While this introduces a bias toward certain types of projects, we argue that it still reflects the current landscape of accessible open-source ML development and offers a diverse set of practical codebases for our analysis.

 \begin{table}[h]
    \centering
    \begin{tabular}{lrr}
        \hline
        \textbf{ML Domain} & \textbf{Project count} & \textbf{Python File count} \\
        \hline
        Audio Processing & 21 & 3273 \\
        Autonomous Gameplay & 36 & 2597 \\
        Image Processing & 34 & 3182 \\
        Natural Language Processing & 7 & 747 \\
        Other ML & 51 & 18192 \\
        Self-Driving Car & 1 & 10 \\
        \hline
    \end{tabular}
    \caption{Projects and file numbers across various domains}
    \label{tab:domain_projects}
\end{table}

\subsection{Non-Machine learning Projects}
To construct our sample of non-ML projects, we follow the same selection process and criteria as for the ML dataset, with two main differences.
First, during the mining step, we used each non-ML specific keyword \textit{without} appending the term \texttt{machine learning}, to avoid ML-related repositories. Second, instead of retrieving only 50 projects per keyword, we collected the top 100 projects per query. This adjustment was necessary since we only used three keywords for non-ML projects, unlike the broader keyword set used for ML projects. The keywords are listed in Table~\ref{tab:ml_domains}.

This initial query resulted in 300 candidate projects, which include both educational and real-world repositories, similar in spirit to the ML dataset. %\LEUSON{You may consider replacing “professional” with “non-educational” or “real-world,” depending on what related studies use.}
Next, we applied the same automatic filtering steps described in Table~\ref{tab:filtering_criteria}. Notably, for the sixth criterion, we inverted the logic: instead of requiring the presence of ML libraries (e.g., PyTorch, TensorFlow), we excluded any project importing such libraries. This step eliminated 106 projects, reducing the pool to 194.

We then conducted a manual validation phase, where two researchers independently reviewed each project to confirm its non-ML nature. This involved checking the README file and source files to ensure the project did not contain ML-related tasks or ML library usage (e.g, import tensorflow as tf). This step refined our dataset to 158 non-ML projects.
Finally, to ensure a fair comparison with our ML dataset, we randomly removed 8 additional projects to reach a balanced sample of \textbf{150 non-ML projects}.

\section{RQ1: \rqone}
\label{sec:RQ1}
\subsection{Motivation}
In their study, Zhang \etal \cite{zhang2023faster} report that performance smells are commonly observed in a variety of Python projects. 
These smells represent inefficient coding practices that can degrade performance. 
However, it remains unclear how these smells are distributed across the different types of projects, particularly between ML and non-ML projects. 
Prior work, such as Zhang's, focuses broadly on detecting and fixing performance smells but does not specify whether there are significant differences between ML and non-ML systems. 
Such a distinction is critical because ML projects often involve unique data processing pipelines \cite{wan2019does}, computationally intensive tasks, and extensive use of specialized libraries \cite{nguyen2019machine}, which may lead to different frequencies of the performance smells evaluated compared to non-ML projects.
Aiming to address this gap, in this research question, we compute and compare the frequency of performance smells across ML and non-ML projects. 

\subsection{Approach}
To answer RQ$_1$, we perform an empirical analysis by computing the frequency of performance smells in our sample of projects.
For that, we use RIdiom, a static code analysis tool developed by Zhang \etal \cite{zhang2022making}, capable of detecting and refactoring non-idiomatic Python code that causes performance smells. 
RIdiom focuses on the nine specific performance smells described in section \ref{sec:background}.

We ran RIdiom on all selected projects in our dataset, both ML and non-ML, using the latest version (0.0.11) of the tool available at the time of the analysis.\footnote{\url{https://pypi.org/project/RefactoringIdioms/}} %\LEUSON{Okay, but what was the version? We need a number}.
For each project, RIdiom was executed once, going through all the Python files and identifying the performance smells along with their location and their fix. %\LEUSON{You have more information reported in the Listing. The information reported here needs to be explained. For example, what does it mean by compli\_code and simple\_code?}.
For each evaluated file, the tool generates a detailed JSON output. As shown in \listref{lst:example_detection}, this JSON contains comprehensive information about the file location of the detected smell, the specific function containing the smell (\textit{me}), and the idiom used to fix the smell identified (\textit{idiom}). The output also includes the problematic code indicated in the \textit{compli\_code} field, the suggested fix provided in the \textit{simple\_code} field, and the relevant line numbers specified in the \textit{lineno} fields.
\begin{lstlisting}[language=Python, caption=Example of the detection produced by RIdiom,label=lst:example_detection,basicstyle=\small\ttfamily] 
{
    "file_path": "../ai_projects/CircuitNet/routability_ir_drop_prediction/train.py",
    "cl": "",
    "me": "train",
    "idiom": "Truth Value Test",
    "compli_code": [
        "iter_num % save_freq == 0"
    ],
    "simple_code": [
        "not iter_num % save_freq"
    ],
    "lineno": [
        [
            [
                161,
                11
            ],
            [
                161,
                36
            ]
        ]
    ],
    "keyno": null}\end{lstlisting}
        
Since some of our analyses are based on the number of lines of code (LOC), before moving on to these analyses, we applied normalization techniques to account for variations in project size. 
This step was essential to ensure that projects with more lines of code do not disproportionately skew the results. 
The normalization was performed by calculating the number of performance smells per KLOC (thousands of lines of code), a common approach used in previous studies \cite{menzies2012local}, which enabled us to make fair comparisons across projects of different sizes.

To guide our analysis, initially, we perform a descriptive analysis of the smells, computing general statistical metrics, like the mean, median, and standard deviation of smell densities per KLOC for both ML and non-ML groups.
This helped us understand the central tendencies and variations in the data. 
We then used box plots to visualize the distribution of smells, which allowed us to identify any outliers or notable differences between the two groups. 

Finally, to statistically compare the occurrence of the smells, we
employed the Mann-Whitney U test, a non-parametric test suitable for comparing two independent groups when the data does not follow a normal distribution. 
Upon checking the normality of our data using the Shapiro-Wilk test, we observed that the smell densities did not follow a normal distribution for both groups, justifying the use of this test.\footnote{\textit{p-value} \textless 0.01, for ML and non-ML sample.} %\LEUSON{Use this pattern from now on.}
% of 0.006 and 2.35e-11 for ML and non-ML, respectively.

\subsection{Results of RQ$_1$}

Table \ref{tab:performance_smells} presents descriptive statistics on the occurrence of performance smells in both machine learning (ML) and non-machine learning (non-ML) projects. The data reveal that the mean density of performance smells is notably higher in ML systems than in non-ML systems, with average values of 1.15 and 0.74, respectively. Additionally, the range of values for ML systems is wider,  suggesting a higher prevalence of performance smells in ML projects.

\begin{table}[h]
    \centering
    \begin{tabular}{|c|c|c|}
        \hline
        \textbf{Statistic} & \textbf{ML Systems} & \textbf{Non-ML Systems} \\
        \hline
        Mean Density of Smells ($\mu$) & 1.15 & 0.71 \\
        \hline
        Standard Deviation ($\sigma$) & 0.62 & 0.43 \\
        \hline
        Minimum & 0 & 0 \\
        \hline
        25th Percentile ($Q_1$) & 0.72 & 0.46 \\
        \hline
        Median ($Q_2$) & 1.05 & 0.68 \\
        \hline
        75th Percentile ($Q_3$) & 1.60 & 0.90 \\
        \hline
        Maximum & 3.27 & 3.43 \\
        \hline
    \end{tabular}
     \caption{Descriptive Statistics of Performance Smells}
    \label{tab:performance_smells}
\end{table}

\begin{figure}
    \centering
    \includegraphics[width=0.8\textwidth]{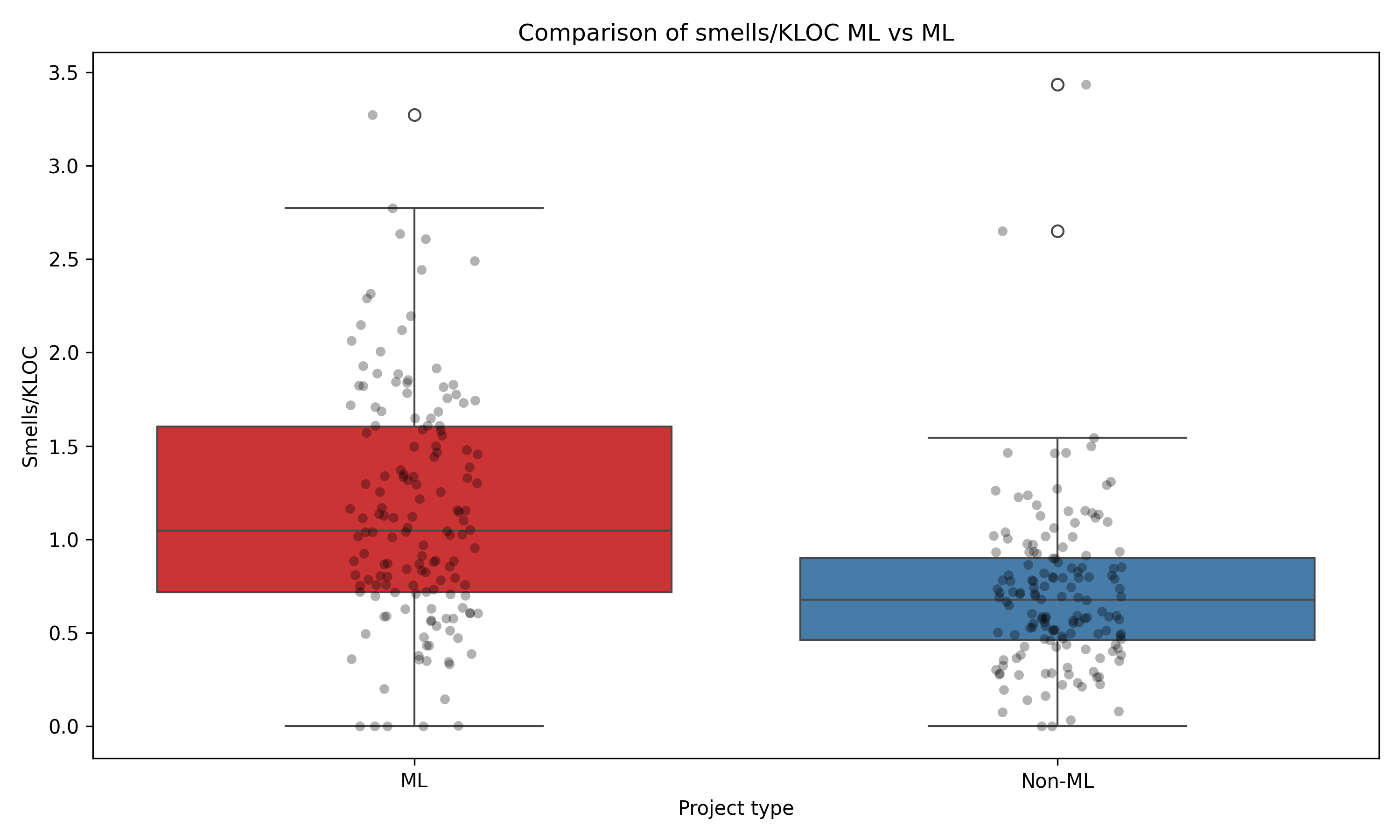}
    \caption{Distribution of performance smells per KLOC
       ML vs Non-ML}
    \label{fig:distributiton_smell_performance_KLOC}
\end{figure}

\begin{figure}
    \centering
    \includegraphics[width=0.8\textwidth]{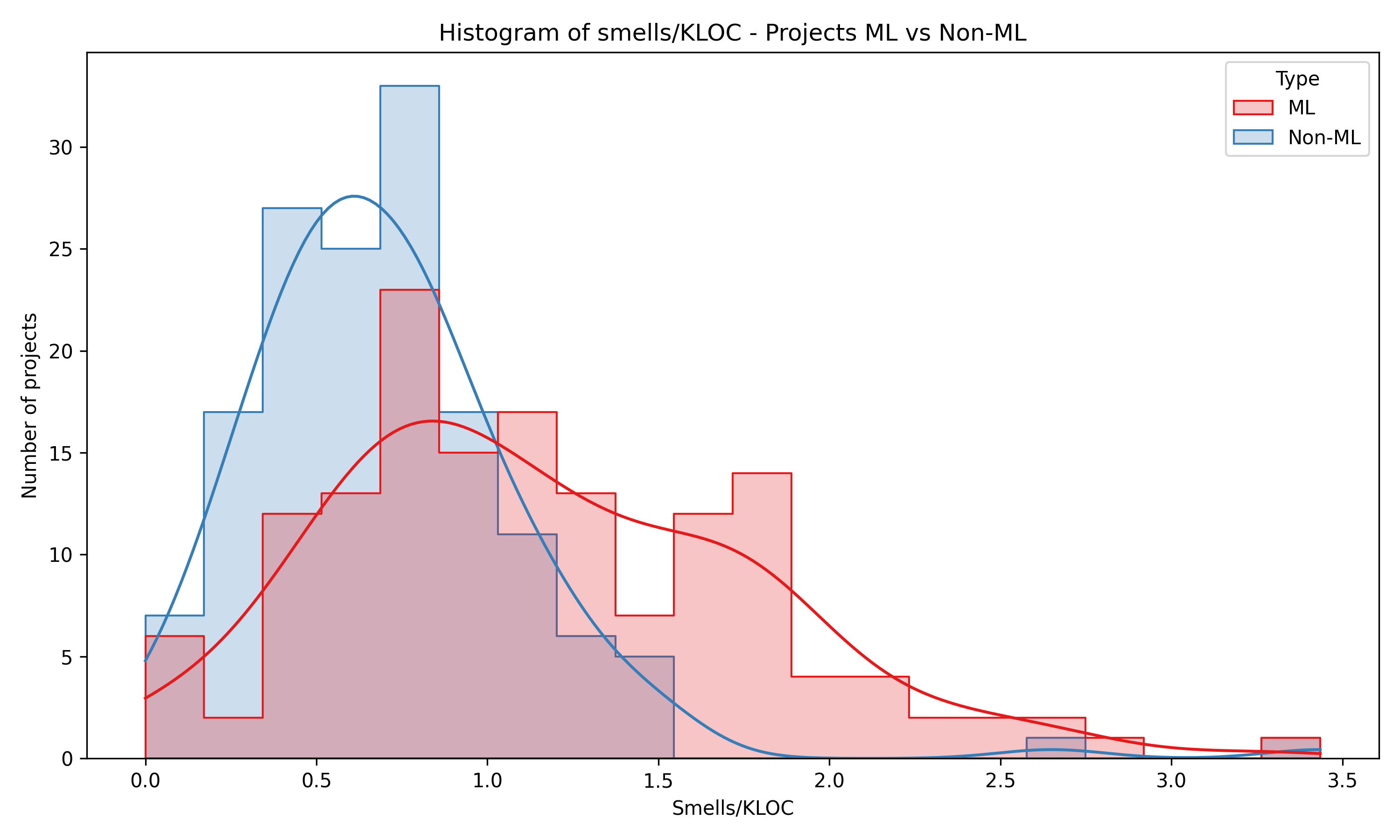}
    \caption{Histogram of smells/KLOC}
    \label{fig:histogram_smell_performance_KLOC}
\end{figure}

Figure \ref{fig:distributiton_smell_performance_KLOC} compares performance smells per 1,000 lines of code (KLOC) between ML and non-ML projects. A significant difference is evident in both the number and the distribution of smells. While ML projects exhibit a higher median of performance smells, approximately 1.05, non-ML projects report a median of 0.68. This highlights a notable disparity in the prevalence of performance smells between the two types of systems.

Several ML projects exhibit smell densities exceeding 2.5 smells/KLOC, which is substantially higher than the typical range observed in non-ML systems. For example, the project Connect-4-Gym-env\footnote{\url{https://github.com/lucasBertola/Connect-4-Gym-env-Reinforcement-learning}}, which implements a reinforcement learning environment for playing Connect-4, exhibits the highest performance smell density in our dataset, with 3.27 smells per KLOC. This project reports 20 distinct files affected by performance smells. Notably, it relies heavily on tight game loops and repeated evaluations of board states, often involving nested control structures and frequent condition testing. 
Additionally, it shows an unusually high density of \textit{Truth Value Test} smells (1.62 per KLOC) and \textit{Chain Compare} (0.20 per KLOC), which may indicate overuse of inefficient complex boolean expressions and inefficient comparisons. %\LEUSON{Don't use " to highlight a term; use textit or something else. } 
Since the project is meant to serve as a testing environment for RL agents, performance might not have been the primary design concern, potentially explaining the accumulation of smells. An example of code exhibiting the presence of performance smells can be the code presented in \listref{lst:example_smell},\footnote{\url{https://github.com/lucasBertola/Connect-4-Gym-env-Reinforcement-learning/blob/main/connect_four_gymnasium/ConnectFourEnv.py}} where we can see that the code does multiple sequential assignements in lines 6-11 instead of using \textit{Assign Multi Targets} construct, inefficient comparison in line 16 instead of using \textit{Chain Compare} and inefficient conditional check in line 26 instead of using \textit{Truth Value Test}. %\LEUSON{You don't have to place the whole function. Use ... to not present everything. Put the code that relates to what you mentioned before, and also, report the tiles of the code you mentioned before. For example, you said additional checks, what lines exactly? You have the numbers. Guide the reader.}\LEUSON{Also, provide the link for the file.}
\begin{lstlisting}[language=Python, caption=ConnectFourEnv.py,label=lst:example_smell,basicstyle=\small\ttfamily]
...
    def __init__(self, opponent=None, render_mode=None, first_player=None, main_player_name='IA'):
        self._opponent = opponent  # Define the opponent
        ...
        
        self.render_mode = render_mode
        self.last_render_time = None
        self.window = None
        self.first_player = first_player
        self.next_player_to_play = 1
        self.main_player_name = main_player_name
        ...

    ...
    def is_action_valid(self, action):
        return action >= self.MIN_INDEX_TO_PLAY and action < self.COLUMNS_COUNT and not self.is_column_full(action)

    def play_action(self, action):
        if not self.is_action_valid(action):
            if self.render_mode == "human":
                print("action_invalid played!")
            self.invalid_move_has_been_played = True
            return 
        
        for i in range(self.ROWS_COUNT - 1, -1, -1):
            if self.board[i, action] == 0:
               ...

\end{lstlisting}

Moreover, ML projects display a wider interquartile range (IQR), indicating greater variability in performance smell occurrence across different projects. 
The whiskers in the box plot further emphasize this disparity, with that of ML projects ranging from 0 to 2.75 smells per KLOC, while non-ML projects range from 0 to 1.25. The presence of outliers, particularly in ML projects, suggests that some projects experience far more performance smells than others. 
Overall, ML projects tend to have a higher concentration of performance smells, accompanied by a more varied distribution.

The histogram representation (Figure \ref{fig:histogram_smell_performance_KLOC}) further highlights these differences. 
Non-ML projects display a more concentrated distribution, with a clear peak between 0.5 and 0.7 smells/KLOC. In contrast, ML projects exhibit a broader, right-skewed distribution with multiple peaks, one around 1 smell/KLOC and another near 2 smells/KLOC, suggesting the presence of potential subgroups exhibiting varying patterns in performance smells.

The Mann-Whitney U test revealed a highly significant difference between the ML and the non-ML systems (p-value $<$ 0.05) %\LEUSON{Check the pattern I mentioned. Usually, you just say p-value \textless 0.05 is enough.}
, confirming that performance smells are statistically more prevalent in ML projects. 
Furthermore, the rank biserial correlation was -0.465, indicating a moderate to large effect size and strengthening the conclusion that ML systems tend to exhibit a higher density of performance smells than non-ML systems.

\begin{resultbox} 
{\bf{RQ$_1$ summary}}: Our findings highlight that performance smells are prevalent in both types of projects, but the density of these smells varies significantly. Specifically, ML projects exhibit a higher mean density of performance smells than non-ML systems, underscoring the importance of addressing these smells during development. 
\end{resultbox}

\section{RQ2: \rqtwo}
\label{sec:RQ2}
\subsection{Motivation}
Among the performance smells analyzed, our goal is to identify those that are more prevalent in machine learning (ML) projects. While Zhang \etal \cite{zhang2023faster} cataloged Python performance smells, they did not explore which smells are most common in ML projects. Similarly, Bhatia \etal \cite{bhatia2023empirical} focused on ML projects but did not specifically examine performance smells. With this research question, we fill this gap by analyzing the distribution and frequency of performance smells in ML projects, helping developers target performance issues that are more likely to arise in ML systems. Furthermore, we compare our results with non-ML projects to understand which smell types are more prevalent in ML systems compared to traditional Python projects.

\subsection{Approach}

Based on the previous reports obtained with RIdiom, for this RQ, we calculate the average frequency of each type of performance smell for both samples of projects (ML and non-ML). We then compare these average frequencies to identify which types of performance smells are more prevalent in ML projects compared to non-ML ones. This comparative analysis allows us to pinpoint smell types that may be characteristic or more frequent in ML systems.
This analysis was conducted at two levels:
\begin{enumerate}
    \item {\bf{File-Level Normalization}}: 
    We first normalized the data based on individual files that exhibit performance issues.  
    This approach allowed us to evaluate the presence of performance smells relative to the number of files containing performance issues in each project type \cite{saboury2017empirical}. %\LEUSON{What do you mean by category? ML or non-ML? That's not a category; it's different samples of projects, no?} 

    \item {\bf{KLOC-Level Normalization}}: Second, we normalized the frequency of smells per 1,000 lines of code (KLOC) to account for differences in project sizes. This enabled us to reassess the distribution of performance smells while ensuring that larger projects did not skew the results \cite{menzies2012local}. 
\end{enumerate}

By analyzing performance smells at both levels, we ensure a comprehensive comparison of their distribution across ML and non-ML projects. 
To assess whether the observed differences are statistically significant, we applied the Mann-Whitney U test to each type of smell, as our dataset does not follow a normal distribution.

\subsection{Results of RQ$_2$}
\begin{figure}
    \centering
    \includegraphics[width=0.8\textwidth]{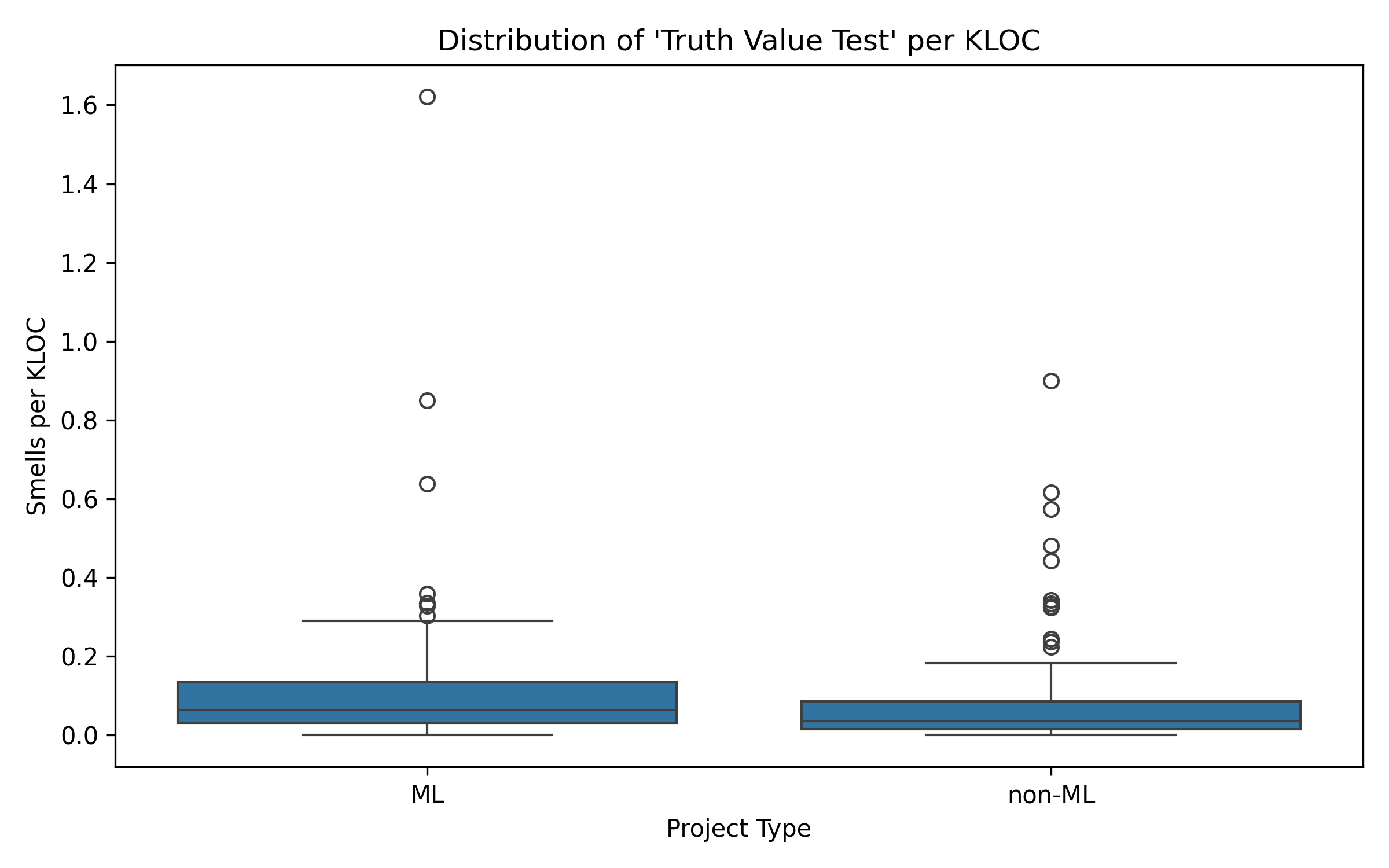}
    \caption{Distribution of \textit{Truth Value Test} per KLOC}
    \label{fig:truth_value_test_per_kloc}
\end{figure}

\begin{figure}
    \centering
    \includegraphics[width=0.8\textwidth]{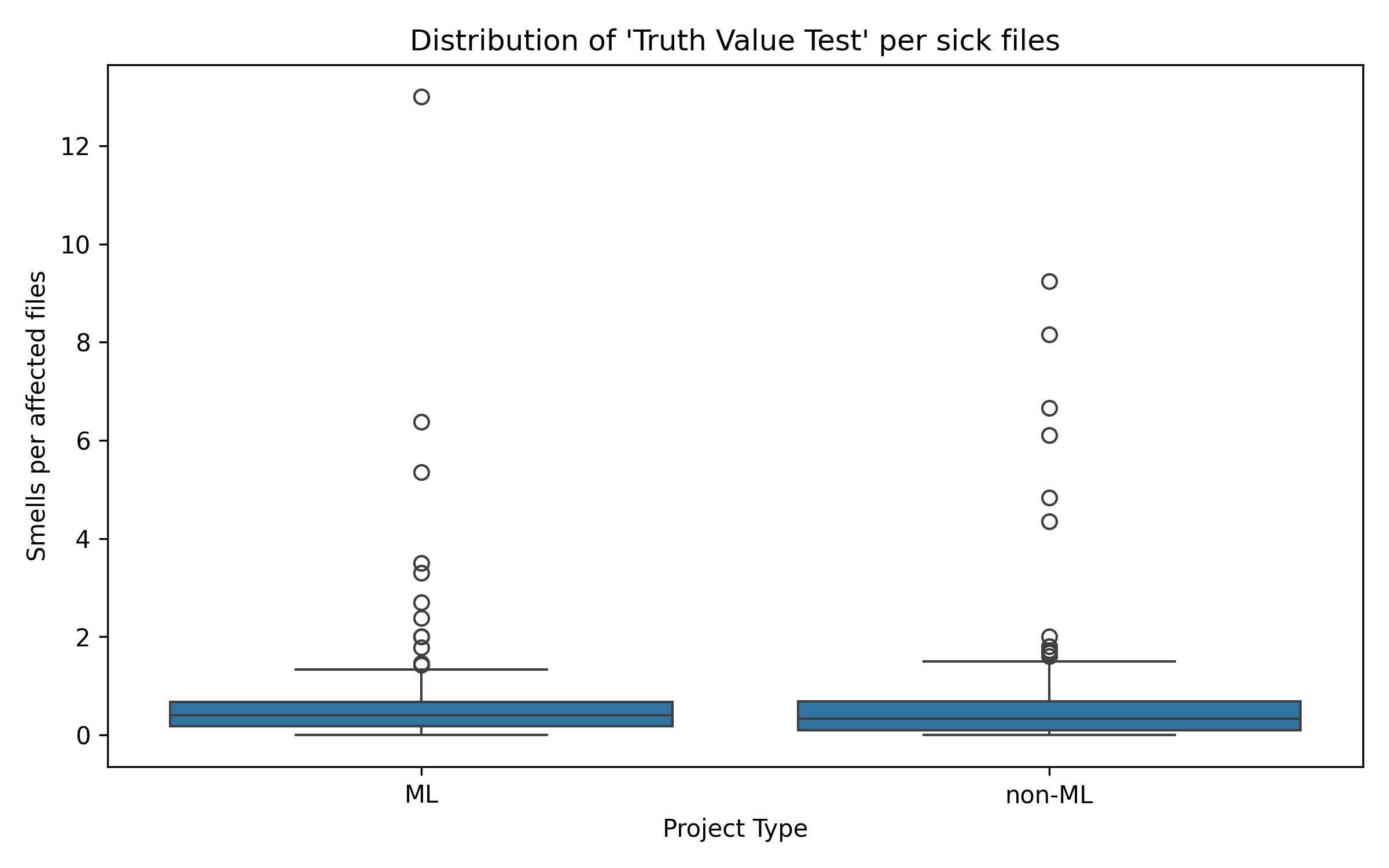}
    \caption{Distribution of \textit{Truth Value Test} per sick files}
    \label{fig:truth_value_test_per_file}
\end{figure}

Table \ref{tab:combined_smells_comparison} compares the occurrence of smells in ML and non-ML projects.
Overall, we observe a higher prevalence of smells in ML projects than in non-ML ones.
However, we observe a statistically significant difference only for some smells: 
\textit{Assign Multi Targets}, \textit{Set Comprehension}, and \textit{Call Star}, which are significant in both normalization methods. 
In contrast, \textit{Chain Compare} and \textit{Truth Value Test} are only significant when normalized by KLOC.
Some smells, such as \textit{List Comprehension}, \textit{Dict Comprehension}, \textit{For Else}, and \textit{For Multi Targets}, do not show any statistically significant difference. To better understand the discrepancy in statistical significance across normalization methods, we analyzed the distribution of the \textit{Truth Value Test} smell using two perspectives: normalization by KLOC and normalization by affected files (i.e., files containing at least one occurrence of the smell). As shown in Figure~\ref{fig:truth_value_test_per_kloc}, when normalized by KLOC, ML projects exhibit a significantly higher density of this smell compared to non-ML projects. This indicates that \textit{Truth Value Test} is more prevalent throughout the codebase of ML projects. However, when the same smell is normalized by the number of affected files (Figure~\ref{fig:truth_value_test_per_file}), the distributions between ML and non-ML projects are quite similar. This suggests that once a file is \textit{infected} by this smell, the intensity (i.e., how many times the smell occurs in that file) does not differ significantly between project types. The detailed p-values for all smell comparisons are reported in Table \ref{tab:merged_mann_whitney_results}, providing a comprehensive statistical foundation for these observations.

\begin{table}[ht]
\centering
\caption{Comparison of Performance Smell Prevalence between ML and Non-ML Projects (per smelly file and per KLOC). %\LEUSON{You call buggy file. Is this term correct? Because the files has no bugs. What about smelly files? Please, check related studies.}
}
\begin{tabular}{lcccc}
\hline
\textbf{Smell Type} & \multicolumn{2}{c}{\textbf{Per Smelly File}} & \multicolumn{2}{c}{\textbf{Per KLOC}} \\
                    & \textbf{ML} & \textbf{Non-ML} & \textbf{ML} & \textbf{Non-ML} \\
\hline
List Comprehension   & 0.1 & 0.07 & 0.017  & 0.0094  \\
Assign Multi Targets & 5.6 & 4.9 & 0.98 & 0.6 \\
Set Comprehension    & 0.002 & 0.004 & 0.00034  & 0.00033  \\
Dict Comprehension   & 0.022 & 0.025 & 0.0039  & 0.0027  \\
Chain Compare        & 0.17 & 0.13 & 0.032  & 0.018  \\
Truth Value Test     & 0.67 & 0.69 & 0.11  & 0.076  \\
Call Star            & 0.06 & 0.032 & 0.0099  & 0.0041  \\
For Else             & 0.006 & 0.005 & 0.001  & 0.00061  \\
For Multi Targets    & 0.037 & 0.025 & 0.0067  & 0.0035  \\
\hline
\end{tabular}
\label{tab:combined_smells_comparison}
\end{table}

\begin{table}[ht]
\centering
\caption{Mann-Whitney U Test Results for Performance Smells (Per Buggy File and Per KLOC). Smells marked with an asterisk (*) indicate statistically significant differences (p $<$ 0.05).}
\label{tab:merged_mann_whitney_results}
\begin{tabular}{|l|c|c|}
\hline
\textbf{Smell Type} & \textbf{p-value (Smelly Files)} & \textbf{p-value (KLOC)} \\
\hline
List Comprehension      & 0.575  & 0.827 \\
Assign Multi Targets*    & 0.03  & 6.24e-12 \\
Set Comprehension*        & 0.0019  & 0.002 \\
Dict Comprehension       & 0.1158  & 0.221 \\
Chain Compare            & 0.5167  & 0.037 \\
Truth Value Test        & 0.5133  & 0.0021 \\
Call Star*                & 0.0053  & 0.0012 \\
For Else                 & 0.2596  & 0.805 \\
For Multi Targets        & 0.9464  & 0.966 \\
\hline
\end{tabular}
\end{table}

The \textit{Assign Multi Targets} smell is notably more prevalent in ML projects, occurring 0.98 times per KLOC compared to 0.6 in non-ML projects (see Table \ref{tab:combined_smells_comparison}). This higher prevalence may be due to the frequent manipulation of large volumes of variables and data in ML projects, which increases the likelihood of such performance smells.
For example, during the training phase of models, several hyperparameters must be configured to find a high-performing model. 
In the AlphaSnake-Zero project\footnote{\url{ https://github.com/Fool-Yang/AlphaSnake-Zero/blob/master/code/utils/alpha_snake_zero_trainer.py}}, we observe many variables being assigned in parallel to control the algorithm parameters as illustrated in \listref{lst:alpha_zero}.

\begin{lstlisting}[language=Python,basicstyle=\small\ttfamily, caption=Example of variable assignments in ML project, label=lst:alpha_zero]
    game_board_height = 11
    game_board_width = 11
    number_of_snakes = 4
    self_play_games = 256
    max_MCTS_depth = 8
    max_MCTS_breadth = 128
    initial_learning_rate = 0.0001
    learning_rate_decay = 0.98
\end{lstlisting}

These multiple assignments illustrate how each parameter directly influences the behaviour and results of the model. Such complexity is inherent in ML projects, where data must be managed and modified frequently throughout the learning process.
In contrast, non-ML projects generally require fewer assignments, as they do not involve the previously mentioned scenarios.  
For instance, in a web development project, most variable assignments are related to storing user inputs from forms or tracking session data.\footnote{\url{https://github.com/Asabeneh/30-Days-Of-Python/blob/master/09_Day_Conditionals/09_conditionals.md}}. %\LEUSON{In this case, the target file is not Python. Is it correct?} \PHILIPPE{not but it contains runnable python code}
The number of assignments is limited to the application's specific requirements, such as saving or processing user data. This way of using assignment variables
is typical of non-ML projects, focusing more on application logic than on tuning multiple parameters. However, this trend cannot be generalized to all non-ML projects; in fact, we observe projects such as graphical ones exhibiting the same need as ML projects to manage visual elements. For example, in the Manim project (specifically the \texttt{Scene} class), assignments are essential for accurately configuring the scene and ensuring smooth rendering.\footnote{\url{https://github.com/3b1b/manim/blob/master/manimlib/scene/scene.py}}

\begin{lstlisting}[language=Python,basicstyle=\small\ttfamily, caption=Example of variable assignments in non-ML project]
    name = request.form['name']
    country = request.form['country']
    city = request.form['city']
    skills = request.form['skills'].split(', ')
    bio = request.form['bio']
    birthyear = request.form['birthyear']
    created_at = datetime.now()
    student = {
        'name': name,
        'country': country,
        'city': city,
        'birthyear': birthyear,
        'skills': skills,
        'bio': bio,
        'created_at': created_at
    }
\end{lstlisting}

Examining the \textit{Truth Value Test} and the \textit{List comprehension} smell types, we observe a high but moderate prevalence in ML projects (Table~\ref{tab:combined_smells_comparison}). The \textit{Truth Value Test} is more frequent in ML projects, likely due to their computational demands. These projects often involve operations on multidimensional arrays, matrices, and tensors that require conditional checks to ensure data integrity and correctness before performing intensive calculations. Similarly, \textit{List comprehension} is commonly used in ML projects because of the frequent reliance on lists to manage datasets. Developers often adopt lists for efficient data handling, yet this can lead to suboptimal looping structures, increasing the likelihood of this smell.
In contrast, non-ML projects typically feature simpler logical structures and handle less data, resulting in fewer occurrences of these performance smells.

For \textit{Set Comprehension}, \textit{Dict Comprehension}, \textit{For Else}, \textit{Call star} and \textit{Chain Compare} smells, the differences between ML and non-ML projects remain minimal. This suggests that their presence is likely influenced more by specific programming needs than by domain-specific conventions or common coding practices.
These smells appear in both ML and non-ML codebases depending on the contextual demands of the task. For instance, dictionary comprehensions may be used in ML to succinctly define model configurations or hyperparameter sets, while in non-ML projects they may serve purposes such as managing cache entries or implementing lookup tables. Similarly, set comprehensions often reflect the need to handle unique elements, be it label sets in ML workflows or deduplication in non-ML scripts. The use of these constructs is thus better understood as a function of problem-specific logic rather than systematic patterns inherent to one domain.

Regarding the \textit{For-else} %\lm{Do you mean the smell Loop-Else?} 
construct, it is relatively rare in Python, as it requires a specific logic pattern where a \textit{break} statement is involved within a loop. 
This pattern is often used in search algorithms to indicate that an item was not found or to signal that all elements were processed without interruption. Its usage is not specific to ML or non-ML projects; rather, it applies universally to any situation where you need to determine if a loop is completed without a break. Due to its specialized nature, its occurrence is not significantly different between ML and non-ML projects.

These findings suggest that the similarities in these smells across both domains are not driven by shared libraries or frameworks like Pandas. Instead, these common smells likely reflect the underlying logic and problem-solving requirements inherent to each project type. Libraries such as Pandas or NumPy, while frequently used in ML, are also employed in non-ML contexts for tasks like data handling, basic calculations, or visualization. However, the presence of smells across both domains appears to be more closely tied to the specific operations that developers implement, which rely on certain patterns regardless of the context. 
This observation indicates that developers in ML and non-ML projects apply these patterns selectively based on the functional demands and the nature of their particular tasks.

\begin{comment}
\begin{resultbox}
{\bf{RQ$_2$ summary}}: Overall, performance smells are more recurrent in ML than Non-ML projects. However, such a difference is statistically significant for specific smells, like \textbf{Assign Multi Targets}, \textbf{Truth Value Test}, and \textbf{List Comprehension}. Such a difference can be attributed to the
manipulation of large datasets by ML projects, which requires extensive variable usage and frequent allocation of data structures.
The other smells, though they are still more prevalent in ML projects, their occurrence does not differ significantly from non-ML projects,
suggesting that other factors might influence such a difference.
\end{resultbox}
\end{comment}

\begin{resultbox}
    {\bf{RQ$_2$ summary}}: Some performance smells are more frequent in ML projects. In particular, \textit{Assign Multi Target}, \textit{\setComp}, and \textit{Call Star} were statistically significant under both normalization methods, while \textit{Chain Compare} and \textit{Truth Value Test} show significance only when normalized by KLOC. These trends may reflect the intensive use of variable assignments and dynamic data structures in ML pipelines. Other smells show no significant difference.
\end{resultbox}

\section{RQ3: \rqthree}
\label{sec:RQ3}

\subsection{Motivation} Cao \etal \cite{cao2022understanding} made an intriguing discovery in their study of performance issues in deep learning systems. Their research, which focused on Deep Learning projects built using TensorFlow and Keras, revealed that performance problems are not evenly distributed across the machine learning pipeline. 
They found that a substantial portion of performance issues emerge in the early stages, particularly during data preprocessing and model building.

Such a finding reveals that different pipeline stages may contribute disproportionately to performance bottlenecks. 
It would, therefore, be valuable to conduct a similar investigation involving different domains of ML projects.
Understanding the distribution of performance smells across the various stages of the machine learning pipeline would allow us to identify critical areas where developers should focus their optimization efforts \cite{foalem2024studying}.

\subsection{Approach}
Since our goal is to evaluate the occurrence of performance smells across different stages of the ML pipeline, we first needed to select a classification that captures the main stages commonly found in practice.
Representations of ML pipelines can vary significantly across sources, projects, and algorithms, with some pipelines being more complex than others. 
For example, some sources adopt a more detailed classification,\footnote{\url{https://quix.io/blog/the-anatomy-of-a-machine-learning-pipeline}} while others adopt a more conservative one.\footnote{\url{https://www.ibm.com/think/topics/machine-learning-pipeline}} %\LEUSON{Footnotes are always placed after the punctuation.}
This diversity highlights the lack of a standard structure, which makes it essential to adopt a consistent and generalizable classification for empirical analysis.
For this study, we adopted the five-stage classification proposed by Foalem \etal~\cite{foalem2024studying}, as it offers a clear and practical framework suited to our analysis. 
In this context, a typical machine learning pipeline comprises the following stages: Data Collection, Data Processing, Model Training, Model Evaluation, and Model Deployment.

Next, we aim to classify the files containing performance smells into the previously defined stages of the ML pipeline.
However, in ML projects, not all files are necessarily related to ML-specific tasks or stages.
For instance, consider the file \texttt{data.py} from the project Machine-Learning-with-Python.\footnote{\url{https://github.com/devAmoghS/Machine-Learning-with-Python/blob/master/k_means_clustering/data.py}}
Based on its filename, one might classify it under the Data Collection stage. Yet, as shown in Listing~\ref{lst:example_file}, this file merely defines a hardcoded array of values, with no actual data loading or preprocessing logic. %\LEUSON{So, what is the conclusion? You present a code, and next, you just change the subject... there is no transition or connection from this paragraph and the next one.}
This example underscores the limitations of relying solely on filenames to infer a file’s role within the ML pipeline. Although heuristic approaches based on naming conventions can enable rapid categorization, they often assume a level of consistency in developer practices that does not hold in real-world projects. Consequently, such heuristics may misclassify files or miss relevant ones altogether. To address this limitation, a more refined approach involves analyzing file content. Drozdova \etal~\cite{drozdova2023code4ml}, for instance, propose using semi-supervised learning to infer the stage of a file based on its code structure and semantics. 
However, such methods face challenges of their own, notably the scarcity of labeled data and the difficulty of handling files that span multiple stages of the ML lifecycle.

\begin{lstlisting}[language=Python, caption=File not related to ML task,label=lst:example_file,basicstyle=\small\ttfamily] 
inputs = [[-14, -5],
 [13, 13],
 [20, 23],
 [-19, -11],
 ...
 [-41, 8],
 [-11, -6],
 [-25, -9],
 [-18, -3]] \end{lstlisting}
 %\LEUSON{You can reduce the contents presented here in the listing. Use ... in the middle, and place only 4 lines before and after. It's enough to transmit the message.}

In our study, we propose a hybrid zero-shot classification approach to address both limitations. 
Our classification approach combines keyword-based pattern matching with a semantic analysis using a pre-trained BERT model. This method does not require any labeled training data and enables multi-label classification of files based on their actual content.
Our approach uses a curated set of stage-specific keywords (see Listing~\ref{lst:classification_keyword}), extracted initially from official documentation and tutorials of popular ML libraries such as Scikit-learn, TensorFlow, and PyTorch. %\LEUSON{Such a threat to validity has to reported.}
These keywords were then manually refined and extended through iterative inspection of real-world ML projects to better reflect common usage patterns. 
For instance, the keyword \texttt{sklearn.datasets} is associated with the Data Collection stage because it relates to loading datasets.

\begin{lstlisting}[language=Python, caption=Stage-specific ML keywords,label=lst:classification_keyword,basicstyle=\small\ttfamily] 
ml_keywords = { 
    "Data Collection": [r"sklearn.datasets", r"load_iris", r"fetch_california_housing",...], 
    
    "Data Processing": [r"StandardScaler", r"MinMaxScaler", r"OneHotEncoder", ...], ... } \end{lstlisting}

\begin{lstlisting}[language=Python, caption=Prompt for zero-shot classification,label=lst:classification,basicstyle=\small\ttfamily] 
stage_descriptions_2 = { 
    "Data Collection": "Code related to gathering datasets, extracting data from APIs, web scraping, or loading data from files or databases. [...]",
    
    "Data Processing": "Code that preprocesses data, including cleaning, normalization, encoding, arrays manipulations, and splitting datasets. [...]", ...}

prompt = "This code is about:" + ", " + "".join([f"{stage}: {description}" for stage, description in stage_descriptions_2.items()])
\end{lstlisting} % I was not sure I don't know if you want me to be more concrete instead of giving the code like this.
%\LEUSON{That's not the prompt. You must report the structure, including how you ask the model, how you structure the different stages. Check related studies and see how they do. }

In parallel, we leverage a pre-trained BERT model, specifically, \texttt{facebook/bart-large-mnli} to perform zero-shot classification.\footnote{\url{https://huggingface.co/facebook/bart-large-mnli}} 
The prompt we use, shown in Listing~\ref{lst:classification}, maps each ML stage to a natural language description that guides the model's predictions.
Each file is first processed by the zero-shot classifier, which computes a confidence score for each ML stage based on the similarity between the file content and the stage descriptions. 
If a file receives a confidence score greater than or equal to 0.9 for one or more reported stages, it is considered to belong to all of these stages. However, for every low score ($<$ 0.9), the system falls back on keyword matching to support the classification. In this case, the file is scanned for stage-specific keywords, and any matches help determine the appropriate label(s). For example, if a file contains both \texttt{load-iris} and \texttt{StandardScaler}, but does not yield high-confidence scores from the classifier, the presence of these keywords would lead to the file being labelled as part of both Data Collection and Data Processing since these terms correspond to those stages, as shown in Listing~\ref{lst:classification_keyword}. %\LEUSON{Guide the reader. Regarding the keyword load-iris, we observe that it is reported in the listing. What about the StandardScaler keyword? So, you can add that information in the previous listing, and here, you refer to it. So, the reader will make a visual association.} %\LEUSON{Please, adopt a standardized way to refer to terms. Sometimes you say \textbf{data collection}, now you report as Data Collection, sometimes with ", now without. Adopt one single approach, and stick to it until the end.}
%\LEUSON{Also, just comment on my comments, instead of removing them. It's hard for me to remember what I asked, and evaluate whether you properly addressed it or not.}
If no high-confidence stage is predicted and no relevant keywords are found, the file is labeled as \textit{unknown}, indicating that it is not clearly associated with any predefined ML stage.
%\LEUSON{Did you discuss how the approach deals when multiple stages are associated with a single file? You must explicitly mention it.} yes in the text above i said this "If a file receives a confidence score greater or equal to 0.9 for one or more stages, it is considered to belong to all of these stages."

To validate our approach's accuracy, we manually evaluated the predictions reported by the model under analysis. Specifically, we randomly sampled 143 files from our dataset and independently labeled each file according to its actual content.
Next, we compute the Cohen’s Kappa score between the annotators, observing a 0.78 score indicating substantial agreement.
We then compared the predicted stages from our approach against this ground truth to compute its classification accuracy and assess inter-annotator reliability.  Table~\ref{tab:classification-accuracy} presents the detailed accuracy results of our classification approach. 
The complete keyword list and stage mappings, along with our implementation code, are included in our replication package \cite{MLvsNO_ML_replication}.

\begin{comment}
In general, the hybrid method achieves an average F1 score of 82\%, an accuracy of 84\%, and a precision of 88\% across all stages of the pipeline, with the Data Collection and Processing stages performing moderately well (F1 = 81\%). The low number of \textit{unknown} classifications (35\%) further suggests that the model captures a substantial portion of the pipeline structure. \LEUSON{You aim to discuss the recall? I think it's a valid discussion to be placed here.}
However, the disparity in performance between stages, as shown in Table \ref{tab:classification-accuracy}, indicates that further refinement is needed, particularly in crafting more precise zero-shot prompts and designing better context-sensitive keyword filters to reduce misclassifications and improve robustness. \LEUSON{I think this paragraph should be placed after the next one, no?}
\end{comment}

In general, the hybrid method achieves an average F1-score of 82\%, an accuracy of 84\%, and a precision of 88\% across all stages of the ML pipeline. The Data Collection and Data Processing stages perform moderately well (F1 = 81\%), while Model Training yields perfect scores across all metrics. The relatively low rate of \textit{unknown} classifications (35\%) further suggests that the model captures a substantial portion of the underlying pipeline structure. However, the disparity in recall across stages especially for Model Evaluation (60\%) and Data Collection (75\%) reveals potential weaknesses in our current keyword and prompt designs. These findings highlight the need for refining zero-shot prompts and enhancing context-aware keyword filters to reduce misclassifications and improve stage coverage. This refinement is essential to increase the robustness of our approach and ensure more consistent classification performance across the pipeline.

\begin{table}[h]
    \centering
    \begin{tabular}{lcccccc}
        \toprule
        & Data Collection & Data Processing & Model Training & Model Evaluation & Model Deployment & Average \\
        \midrule
        Precision & 0.90 & 0.95 & 1.00 & 0.86 & 0.73 & 0.888 \\
        Recall & 0.75 & 0.70 & 1.00 & 0.60 & 0.86 & 0.782 \\
        F1-score & 0.81 & 0.81 & 1.00 & 0.71 & 0.79 & 0.824 \\
        Accuracy & 0.84 & 0.73 & 1.00 & 0.80 & 0.84 & 0.842 \\
        \bottomrule
    \end{tabular}
    \caption{Performance of the classification model across all Machine learning stages.}
    \label{tab:classification-accuracy}
\end{table}

%\LEUSON{For the table, try to break the terms composed of two words. For example, instead of Data Collection, you can use Data breakline Collection.} don't know how to do it will address it later

\subsection{Results of RQ$_3$}
\begin{figure}
    \centering
    \includegraphics[width=0.8\textwidth]{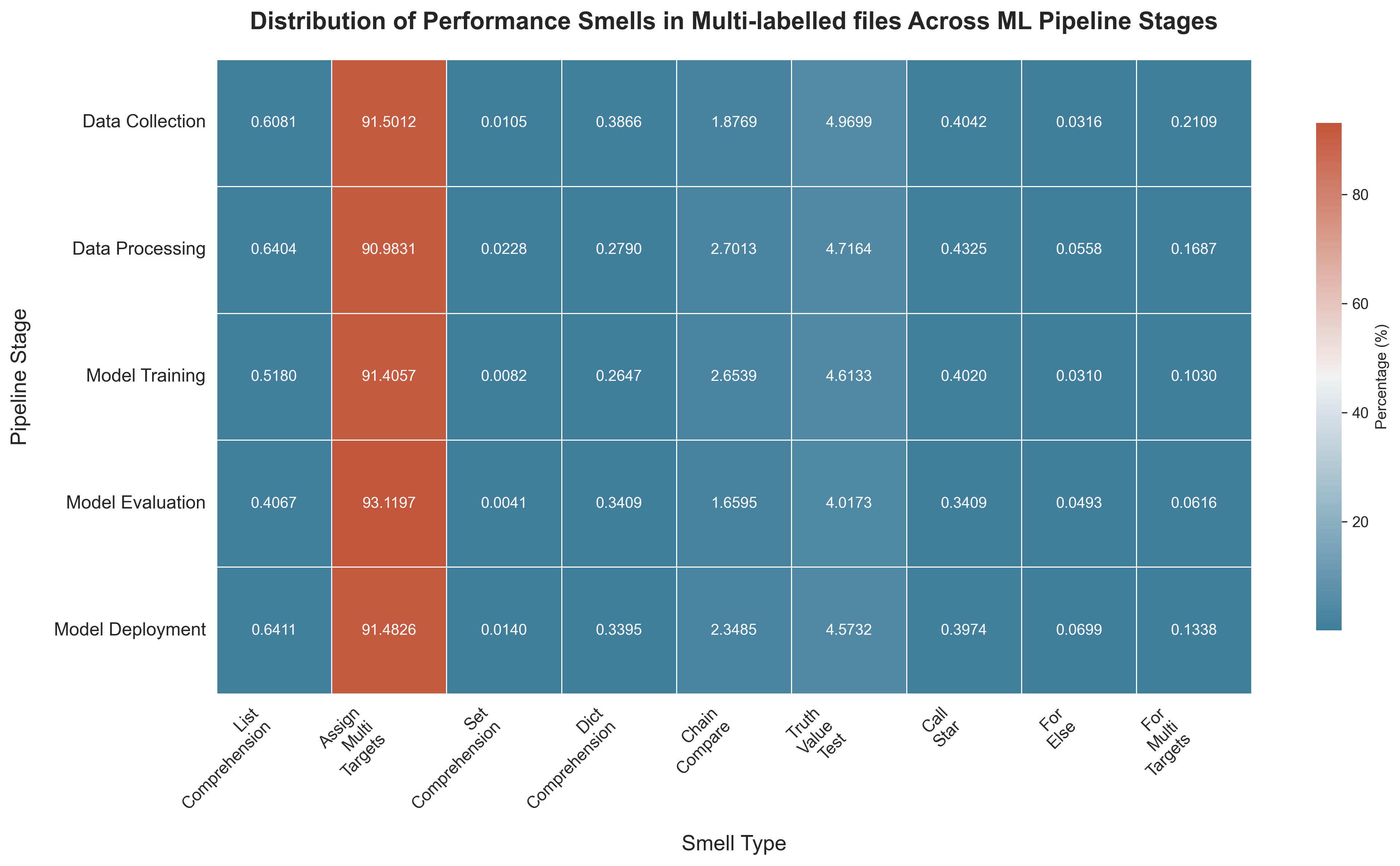}
    \caption{Distribution of Performance Smells in Multi-labelled files Across ML Pipeline Stages (\%)}
    \label{fig:multi_label}
\end{figure}

\begin{figure}
    \centering
    \includegraphics[width=0.8\textwidth]{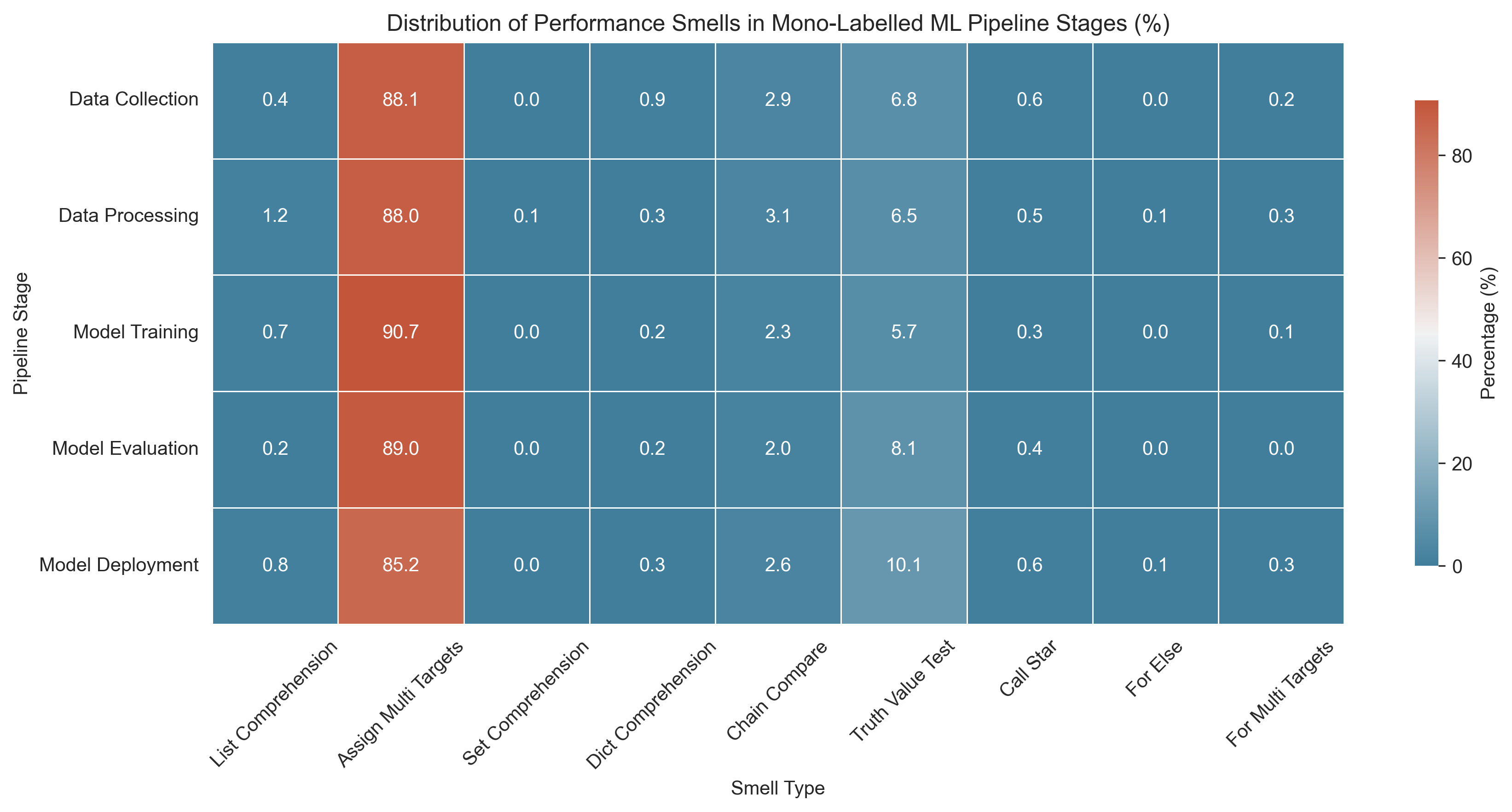}
    \caption{Distribution of Performance Smells in Mono-Labelled Files ML Pipeline Stages (\%)}
    \label{fig:mono_label}
\end{figure}

%\LEUSON{Before you start your results, guide the user. If you start with an overall analysis, say that you start with that. Then, inform the reader you're going into details regarding the stages analyzed here. Right now, you have 5 pages of results, with no transition, and very confusing. Use subsections to structure the results.}
To answer this RQ, we first provide an overall analysis of the performance smells distribution (Figures~\ref{fig:multi_label} and~\ref{fig:mono_label}); we structure our analysis by initially examining the multi-labeled dataset representation, where files can be associated with multiple stages, and then complement it with findings from the mono-labeled subset, where each file is assigned a single dominant pipeline stage.
Then, we provide a detailed examination of each ML pipeline stage. 

%Figures~\ref{fig:multi_label} and~\ref{fig:mono_label} present heatmaps showing the distribution of performance smells across the ML pipeline stages. In this section, we first provide an overall analysis of the performance smells distribution, followed by a detailed examination of each ML pipeline stage. We structure our analysis by first examining the multi-labeled dataset representation, where files can be associated with multiple stages, and then complement it with findings from the mono-labeled subset, where each file is assigned a single dominant pipeline stage.

\subsubsection{Multi-labeled Stage Analysis}

In this analysis, we focus on the files that present performance smells and are associated with multiple stages. 
Such an analysis reflects the reality of ML files that often span several stages (e.g., Data Collection and Processing). The heatmap representing performance smells in multi-labeled files reveals significant patterns across ML pipeline stages (Figure \ref{fig:multi_label}).
Data Processing emerges as the most affected stage when considering the diversity of performance smells. 
While \textit{Assign Multi Targets} is prevalent across all stages ($>$ 90\%), Data Processing shows the most distributed pattern of performance smells, with notable percentages in \textit{List Comprehension}, \textit{Chain Compare}, \textit{Truth Value Test}, and \textit{For Multi Targets}. 
Such findings suggest that Data Processing involves more complex code structures prone to different performance problems.

The diversity of smells in the Data Processing stage likely stems from the variety of tasks performed during this phase, including data cleaning, transformation, and feature engineering. These operations often require complex manipulations, leading to inefficient code patterns if not carefully implemented. 
The stage's position as a bridge between raw Data Collection and Model Training may also contribute to its complexity, as it must handle both unstructured inputs and prepare structured outputs suitable for training algorithms \cite{kuchnik2022plumber}.
%\LEUSON{You informed the reader that you will start with the multi-labelled heatmap. Okay, do it. Present the results. When you're done, you bring into the discussion the possibility of threats and motivate about the mono-labelled approach. }

However, such analysis may introduce an over-coverage effect, where certain smells are attributed to multiple stages, possibly inflating their presence; we discuss in detail such a threat in Section~\ref{sec:threats}.
To mitigate this issue, we complement our analysis with a mono-labelled subset of the dataset, where each file is assigned a single dominant pipeline stage. 
Such a decision allows for a more specific attribution of smells and facilitates the identification of patterns between code smells and pipeline responsibilities.

\subsubsection{Mono-labeled Pipeline Analysis}
When examining the mono-labeled heatmap (Figure \ref{fig:mono_label}), we observe similar patterns compared to our previous analysis, but with distinct characteristics. 
First, Data Processing still stands out as the most affected stage, confirming our findings from the multi-labeled analysis. 
This stage registers the highest percentage for \textit{List Comprehension} (1.2\%) and \textit{Chain Compare} (3.1\%), while also showing measurable values for all other smell types. The consistency of these findings across both labeling approaches suggests that Data Processing genuinely contains more diverse performance issues, likely due to the complex data transformations and cleaning operations that occur in this stage.

Interestingly, the Model Deployment stage appears to exhibit the highest percentage of \textit{Truth Value Test} smells (10.1\%). 
However, given that the classifier's precision for this stage is 73\% as we can see in Table \ref{tab:classification-accuracy}, this observation may be partially influenced by misclassification. 
To account for this observation, we further inspected high-confidence samples from this class, which revealed a stronger and more consistent presence of \textit{Assign Multi Target} smells instead. 
Such a finding suggests that while conditional checking might be common in deployment-related code, possibly due to error handling and system safeguards, the true distribution of smell types warrants more careful investigation with a more precise classifier. 
These preliminary findings nonetheless highlight the potential impact of certain performance smells on runtime performance in production environments, suggesting areas for targeted future optimization.

Model Evaluation and Model Training stages share similar performance smell profiles in both labeling approaches, suggesting that the code patterns used in model creation and assessment may be functionally related. This similarity makes sense given that these stages often involve iterative processes of model training, testing, and validation. The slightly higher prevalence of \textit{Chain Compare} smells in Model Training compared to Model Evaluation (2.5\% vs. 2.0\%) might reflect more complex parameter condition checking during the training process.

\subsubsection{Cross-cutting Patterns and Implications} 
%\LEUSON{I wonder whether this section could be placed in Section 7.} I prefer to keep this section here as it summarizes empirical patterns rather than offering prescriptive guidelines.
Overall, our analyses reveal consistent patterns that provide valuable insights into ML pipeline quality characteristics. 
The overwhelming dominance of \textit{Assign Multi Targets} across all pipeline stages ($>$ 85\% in all cases) suggests that Python's multiple assignment feature is fundamental to ML development, regardless of the pipeline stage. 
This prevalence likely reflects the data-centric nature of ML tasks, where multiple variables are often assigned in multiple lines of statements instead of a single one.

The consistent secondary prevalence of \textit{Truth Value Test} and \textit{Chain Compare} smells across stages indicates common patterns in conditional logic throughout ML pipelines. These smells often indicate inefficient boolean expressions or comparison operations, which may impact performance in data-intensive ML applications. The presence of these smells across all stages suggests that optimization techniques targeting these specific code patterns could yield performance improvements throughout the entire ML workflow.
On the other hand, the relative rarity of \textit{Set Comprehension} and \textit{For-Else} smells likely reflects the low presence of code patterns that could benefit from these constructs. This observation aligns with our findings in Section~\ref{sec:RQ2}, which noted that such constructs are typically used for specific tasks only.

%the less common usage of these language features in ML development, as opposed to traditional Python applications. This finding aligns with typical ML coding practices, where list comprehensions are more commonly used than set comprehensions, and traditional loop structures are preferred over the more specialized \textit{For-else} construct.

%\LEUSON{That's a common mistake. You're replicating the numbers from the figure; the reader can read and understand your figure. Instead, focus on providing the insight behind. I would suggest discussing the stages first, as that's the focus here. Of course, you can bring up some discussion about the smell types, but here the focus is on the ML pipeline stages.} \LEUSON{Adjust on the usage of ". Use another way to highlight the text; I prefer textit{} for example.}

%\LEUSON{It's not to have bullet points pointing out how they share the same ideas. Please discuss your findings. Reasons for such an observation.  Instead, you're reporting smell type frequencies. Again, the reader can see it from the charts. So, remove it or bring it as a paragraph, if necessary.}

When comparing both analyses, we observe that general trends remain consistent, but the mono-labelled version provides sharper contrasts and reduces noise from cross-stage contamination, giving us confidence that our findings represent genuine patterns rather than artifacts of our labeling approach. This consistency across different analytical approaches strengthens the validity of our results and suggests that the identified patterns of performance smells are inherent to the functional requirements of each ML pipeline stage rather than being coincidental. 
%\LEUSON{Here, you point to the evidence that the initial threat is minimized, considering both approaches share the same findings. That's the goal.}

\begin{comment}
\PHILIPPE{Figures \ref{fig:multi_label} and \ref{fig:mono_label} present heatmaps illustrating the distribution of performance smells across different stages of the ML pipeline. To analyze the variation of performance smells, we used a mono-labelled subset of our dataset to mitigate the over-coverage effect observed in multi-labelled files (cf. Section \ref{sec:threats}). \LEUSON{What do you mean here? You can have both approaches. First, present one and discuss. Then, you motivate the second, present, and discuss. Right now, it's confusing. I'd suggest starting with the multi-labelled files approach. And then, you motivate the need for a one-labeled approach.
After, you can compare whether they share the same conclusions.}
The resulting heatmap provides a clearer attribution of performance smells to specific pipeline stages. 
To understand why certain smells are more prevalent in certain stages, we performed a manual inspection of 10 randomly selected files for each stage. \LEUSON{That's clearly methodology; it should not be here.} 
\end{comment}
\subsubsection{Pipeline Stage Analysis}
Based on the previous results, here, we analyze the ML pipelines stages to better understand the nature of the operations being carried out, while reflecting them with the occurrence of performance smells.

\subsubsection*{Data Collection} 
This stage encompasses the extraction of data from external sources such as datasets, APIs, and platforms like Kaggle~\cite{8862913}. 
During this process, developers implement validation mechanisms to verify critical data properties (e.g., dimensional consistency, format compliance, or completeness), which explains the prevalence of code patterns that could benefit from \textit{Truth Value Test} optimizations.
This stage also exhibits a high frequency of variable initialization patterns. Since ML pipelines typically require multiple data structures to simultaneously manage different components of the incoming data (e.g., features, labels, and metadata), developers frequently create sequences of assignments that could be more efficiently expressed using the \textit{Assign Multi Targets} idiom.

Moving forward, consider the example in the \texttt{GroupedBatchSampler} class from the deep-learning-for-image-processing project\footnote{\url{https://github.com/WZMIAOMIAO/deep-learning-for-image-processing/blob/master/pytorch_keypoint/HRNet/train_utils/group_by_aspect_ratio.py}}. As shown in \listref{lst:example_dc}, sequential variable initializations occur in lines 7-9 and 12-13, typifying the structure preparation phase where counters and buffers are established before actual data acquisition begins. Additionally, the conditional checks in lines 23-24 govern the sampling logic by verifying data availability within groups, ensuring complete data retrieval.
These code segments contain inefficiencies addressable through \textit{Assign Multi Targets} and \textit{Truth Value Test} optimizations. The performance consequences of these inefficiencies are significant, as the \texttt{\_\_iter\_\_} method may be called repeatedly throughout training cycles. Suboptimal implementations generate substantial latency in the data pipeline, creating bottlenecks that affect the entire training workflow. By optimizing these patterns, developers can enhance data loading efficiency and improve the ML training loop's responsiveness, benefits that become especially critical in time-sensitive applications or environments with limited computational resources.

\begin{lstlisting}[language=Python, caption=Example of operations done in the Data Collection stage,label=lst:example_dc,basicstyle=\small\ttfamily]
def __init__(self, sampler, group_ids, batch_size):
        if not isinstance(sampler, Sampler):
            raise ValueError(
                "sampler should be an instance of "
                "torch.utils.data.Sampler, but got sampler={}".format(sampler)
            )
        self.sampler = sampler
        self.group_ids = group_ids
        self.batch_size = batch_size
    
def __iter__(self):
    buffer_per_group = defaultdict(list)
    samples_per_group = defaultdict(list)
    ...

    if num_remaining > 0:
            # for the remaining batches, take first the buffers with largest number
            # of elements
            for group_id, _ in sorted(buffer_per_group.items(),
                                      key=lambda x: len(x[1]), reverse=True):
                ...
                num_remaining -= 1
                if num_remaining == 0:
                    break
        assert num_remaining == 0
 \end{lstlisting}

%\LEUSON{One example is enough; since you have two, and they share the same target problem, provide both code snippets and discuss them.} where multiple data structures are populated and validated using assert and if checks as illustrated in \listref{lst:example_dc}. \LEUSON{The idea to bring code here is to help you report your results. Right, now you just present the code, and there is transition with the text. Explain the code smell in the code. Then, discuss the possible impact in this stage. Point out the lines. Reduce the code if necessary. Check related studies, please, and see how they use code. Check this paper: https://arxiv.org/pdf/2402.08801, pages 15-16. I'll revise it when you're done with this overall section.}

\subsubsection*{Data Processing}
This stage involves critical tasks such as data cleaning, transformation, and augmentation. These operations extensively manipulate data structures, like reshaping tensors, converting formats (e.g., from NumPy arrays to PyTorch tensors), and applying element-wise transformations. 
Consequently, we observe a higher prevalence of patterns that could benefit from \textit{Truth Value Test}, \textit{Chain Compare}, and \textit{Assign Multi Targets} optimizations, which can significantly impact performance.

Conditional validations are particularly frequent when verifying tensor shapes, checking value ranges, or ensuring consistency of transformed data. The complexity of these operations often leads to suboptimal conditional expressions that could be refactored for both clarity and efficiency. We identified a representative example in the \texttt{DouDizhuRuleAgentV1} %\LEUSON{In the previous example, you used texttt, now you used textit. Standardize and stick to it. For code, I'd prefer texttt} 
class (specifically the \texttt{step} and \texttt{pick\_chain} methods) from the rlCard project\footnote{\url{https://github.com/datamllab/rlcard/blob/master/rlcard/models/doudizhu_rule_models.py}}. As shown in \listref{lst:example_dp}, line 6 contains an inefficient check that could be refactored to \texttt{if not len(trace)}, while line 22 could be simplified to \texttt{if min\_count}. %\LEUSON{The informed lines are not associated with code in the listing. Also, when reporting to code, use the texttt} 
These refactorings not only reduce verbosity but also improve execution performance by eliminating redundant comparisons and leveraging Python's internal truthiness evaluation.

Furthermore, the same line 6 of the code could benefit from \textit{Chain Compare} optimization, allowing for the combination of related boolean expressions into a more efficient form such as \texttt{len(trace) $>=$ 3 and trace[-1][1] == 'pass' == trace[-2][1]}. This chained comparison minimizes the number of separate conditional evaluations and enables faster short-circuit evaluation paths at runtime \cite{zhang2023faster}. Such optimizations are particularly valuable in Data Processing operations that may be applied repeatedly across large datasets, where even small inefficiencies can compound into significant performance penalties \cite{huang2015empirical}. %\LEUSON{Here, you report Data Processing. Add a new command for each ML stage, and then we'll have a standardized output.} I don't understand this comment at all yes i report Data Processing my output is the same across the file D and P each time no ?

\begin{lstlisting}[language=Python, caption=Example of operations done in the Data Processing stage,label=lst:example_dp,basicstyle=\small\ttfamily]
def step(self, state):
        ...
        state = state['raw_obs']
        trace = state['trace']
        # the rule of leading round
        if len(trace) == 0 or (len(trace) >= 3 and trace[-1][1] == 'pass' and trace[-2][1] == 'pass'):
            comb = self.combine_cards(state['current_hand'])
            min_card = state['current_hand'][0]
            for _, actions in comb.items():
            ....
            return np.random.choice(state['actions'])


...
def pick_chain(hand_list, count):
        chains = []
        ...
        for index, chain in enumerate(chain_list):
            if len(chain) > 0:
                if len(chain) >= 5:
                    start = index + add
                    min_count = int(min(chain)) // count
                    if min_count != 0:
                        str_chain = ''
                        ...

    
 \end{lstlisting}

\subsubsection*{Model Training}
This process involves defining and connecting multiple layers of neural networks, each of which requires its own configuration and initialization steps. 
As model complexity increases, the need to track various components such as %\LEUSON{Incomplete sentence... missing verb}: 
multiple layer instances, training hyperparameters (e.g., learning rate, batch size, number of epochs), and training results becomes significantly more pronounced. 
Such operations lead to frequent sequences of individual assignments that could be more elegantly handled using the \textit{Assign Multi Targets} idiom. Additionally, developers implement verification steps to ensure layers are correctly initialized and behave as expected, resulting in numerous conditional checks throughout the code.

The accel-brain-code repository\footnote{\url{https://github.com/accel-brain/accel-brain-code/blob/master/Accel-Brain-Base/accelbrainbase/controllablemodel/_mxnet/adversarial_ssda_controller.py}} provides an illustrative example, where multiple parameters are sequentially assigned to variables (lines 2-10), and conditional checks are implemented to ensure robustness, as shown in \listref{lst:example_mt}.
The code also employs verbose conditional checks, as evident in lines 13, 16, and 19. These instances could benefit substantially from the \textit{Chain Compare} feature, which would reduce redundant comparison operations and improve code readability \cite{zhang2023faster}. Similarly, lines 24 and 27 contain modulo-based checks that control when certain operations are executed. These could be optimized using the \textit{Truth Value Test} idiom, enhancing both readability and computational efficiency \cite{zhang2023faster}.
The persistence of these code smells introduces redundant operations that may significantly increase model training time if not addressed.
\begin{lstlisting}[language=Python, caption=Example of operations done in the Model Training stage,label=lst:example_mt,basicstyle=\small\ttfamily]
...
        self.__learning_rate = learning_rate
        self.__learning_attenuate_rate = learning_attenuate_rate
        self.__attenuate_epoch = attenuate_epoch
        self.__ctx = ctx
        self.__loss_list = []
        self.__acc_list = []
        self.__target_domain_arr = None
        self.__tol = tol
        self.__est = est
...

        if isinstance(encoder, ConvolutionalNeuralNetworks) is False:
            raise TypeError("The type of `encoder` must be `ConvolutionalNeuralNetworks`.")

        if isinstance(classifier, NeuralNetworks) is False and isinstance(classifier, ConvolutionalNeuralNetworks) is False:
            raise TypeError("The type of `classifier` must be `NeuralNetworks` or `ConvolutionalNeuralNetworks`.")
        
        if isinstance(pretext_task_model, NeuralNetworks) is False and isinstance(pretext_task_model, ConvolutionalNeuralNetworks) is False:
            raise TypeError("The type of `pretext_task_model` must be `NeuralNetworks` or `ConvolutionalNeuralNetworks`.")
...
                if est_flag is False:
                    if tol_flag is True:
                        if ((epoch + 1) % 100 == 0):
                           ...

                    if (iter_n+1) % int(iteratable_data.iter_n / iteratable_data.epochs) == 0:
                    ...\end{lstlisting}
%\LEUSON{I think you can reduce the listings. for example, here you discuss around 8 lines, but your file has 35. So, most of the lines are not even necessary. As I recommended before, use ... etc.}

\subsubsection*{Model Evaluation} 
This stage is responsible for assessing model accuracy using different metrics such as the F1 score and confusion matrices. 
As such, it involves storing intermediate scores, predictions, and statistical results in variables, giving opportunities for optimization patterns like \textit{Assign Multi Targets}. 
In practice, developers often compare expected and actual outcomes across various inputs or thresholds, resulting in repetitive conditional logic. These patterns present opportunities for more efficient constructs like \textit{Truth Value Test} and \textit{Chain Compare}.

The Overcooked-AI project illustrates these patterns well.\footnote{\url{https://github.com/bic4907/Overcooked-AI/blob/main/train_online.py}} In lines 7–10 of \listref{lst:example_me}, several variables are initialized separately. This approach, common in evaluation code where test environments are reset and metrics reinitialized between runs, could be made more efficient by using \textit{Assign Multi Targets}, improving runtime performance.\footnote{\url{https://stackoverflow.com/questions/22278695/python-multiple-assignment-vs-individual-assignment-speed}} %\cite{zhang2023faster}. \LEUSON{You rely a lot on this reference 40. There is no other reference to support your claims?}
Similarly, the \textit{Truth Value Test} idiom can streamline frequent conditional checks. For instance, line 20 of \listref{lst:example_me} demonstrates a case where an unnecessary comparison could be eliminated, providing a noticeable benefit in scenarios where evaluations are executed thousands of times.
Additionally, \textit{Chain Compare} enhances efficiency by reducing redundant variable accesses and logical operations. For instance, line 17 could be rewritten using this idiom for better efficiency. 

These micro-optimizations are particularly valuable when evaluating large models with frequent checkpoint validations. In compute-intensive contexts such as reinforcement learning, where training and evaluation compete for system resources, these enhancements can substantially reduce overhead.

\begin{lstlisting}[language=Python, caption=Example of operations done in the Model Evaluation stage,label=lst:example_me,basicstyle=\small\ttfamily]
...
def evaluate(self):
        average_episode_reward = 0

        self.video_recorder.init(enabled=True)
        for episode in range(self.cfg.num_eval_episodes):
            obs = self.env.reset()
            episode_step = 0
            done = False
            episode_reward = 0
            ...
            average_episode_reward += episode_reward
        self.video_recorder.save(f'{self.step}.mp4')

 def run(self):
 ...
    if self.step >= self.cfg.num_seed_steps and self.step >= self.agent.batch_size:
        self.agent.update(self.replay_buffer, self.logger, self.step)
 ...
   if self.step % 5e4 == 0 and self.save_replay_buffer:
                self.replay_buffer.save(self.work_dir, self.step - 1)
...
 \end{lstlisting}

\subsubsection*{Model Deployment}
This stage generally involves loading pre-trained models, configuring their runtime environment, feeding them live or simulated input data, and occasionally re-evaluating their outputs. In the analyzed code samples, which were filtered to ensure high confidence in the classification, we observed recurring patterns such as model loading and configuration setups. These operations often rely on structured assignments and conditional checks to ensure runtime robustness. However, if not managed properly, these patterns can lead to performance bottlenecks or reliability issues in production environments.

Initially, we observed \textit{Truth Value Test} smells in this stage, but a closer inspection showed that \textit{Assign Multi Target} better characterizes the actual deployment code.
This suggests that using \textit{Assign Multi Targets} may be beneficial here, as sequential assignments are a common pattern.
A representative example appears in the Nkululeko repository\footnote{\url{https://github.com/felixbur/nkululeko/blob/main/nkululeko/runmanager.py}}: lines 6–7, 13–14, and 25–26 in \listref{lst:example_md} could be improved by using \textit{Assign Multi Target} to increase clarity and possibly performance.
 %\LEUSON{That's confusing to me... even when reading from the results, I did not get the intuition behind it.}

\begin{lstlisting}[language=Python, caption=Example of operations done in the Model Deployment stage,label=lst:example_md,basicstyle=\small\ttfamily]
    ....
    def __init__(self, df_train, df_test, feats_train,
                 feats_test, dev_x=None, dev_y=None):
                 ...
        self.df_dev, self.feats_dev = dev_x, dev_y
        self.util = Util("runmanager")
        self.target = glob_conf.config["DATA"]["target"]
        self.split3 = eval(self.util.config_val("EXP", "traindevtest", "False"))
        ....

    def do_runs(self):
        """Start the runs."""
        self.best_results = []  # keep the best result per run
        self.last_epochs = []  # keep the epoch
        ...

    ....
     def search_best_result(self, reports, order):
        best_r = Reporter([], [], None, 0, 0)
        if order == "ascending":
            best_result = 0
            for r in reports:
                res = r.result.test
                if res > best_result:
                    best_result = res
                    best_r = r
                ....
        return best_r

 \end{lstlisting}

%\LEUSON{If I remember well, one of the reviewers asked for analysis considering the different types of projects. We have to present something about it, considering we classified projects in the beginning; so, if such a piece of information is not used, why did we categorize the selected projects?} \PHILIPPE{I removed that part as I didn't do a study per category. Retrieving projects across multiple categories (server, library, etc ...) helps to be broad in the study.}

\begin{resultbox}    
{\bf RQ$_3$ summary}: Our results show that Data Processing is the most affected ML pipeline stage, with Data Collection also showing a high concentration of performance smells. This supports prior findings that early stages are particularly smell-prone. Interestingly, smells also appear in Model Deployment, confirming that inefficiencies can arise throughout the pipeline. %Finally, analyzing mono-labelled files reveals that smell types vary by stage, highlighting the need for stage-specific mitigation strategies. %\LEUSON{Reduce it. it's a summary. Here, you must exercise your capability to summarize the main findings of your results. You can even check the paper I shared here before: the longest summary box has 6 lines, i'd say.}
\end{resultbox}

\section{Guidelines and Implications}
\label{sec:research_guidelines}

Our study has several practical and research implications that can guide developers, tool designers, and the research community. In what follows, we outline key insights and actionable recommendations based on our findings, organized by stakeholder group.

\subsection{Implications for ML Developers}
\subsubsection*{Prioritize Optimization in Data Processing Stages}
We observed that the Data Processing stage is consistently the most affected by performance smells. This stage is central to transforming and preparing data, and inefficiencies here tend to propagate throughout the pipeline, degrading the performance of subsequent stages \cite{huang2015empirical}. Developers are therefore encouraged to prioritize this stage during optimization efforts. Specifically, careful smell detection and refactoring during data preprocessing phases, as well as the use of efficient constructs (e.g., optimized loop, list, comparison operations), can lead to better resource usage. These recommendations align with green software engineering goals, as optimizing this stage can significantly reduce energy consumption in ML workflows \cite{kuchnik2022plumber}.

\subsubsection*{Adopt Stage-Aware Optimization Strategies}
Our results indicate that performance smells are not uniformly distributed but instead reflect the functional nature of each pipeline stage. For instance, the Data Processing stage often involves inefficient tensor manipulations, while the Model Deployment stage is more prone to inefficient assignement operations. This variability implies that smell detection strategies should be context-aware, adapting their logic and recommendations according to the specific responsibilities of each stage. Rather than applying generic detection rules, tools should incorporate domain knowledge of ML practices to identify performance-critical smells and guide developers toward relevant optimizations. Doing so ensures that the focus remains not only on code maintainability, but also on runtime performance and resource efficiency.

\subsection{Implications for Tool Designers}
\subsubsection*{Develop Stage-Aware Smell Detection Tools}
Based on our findings, smell detection strategies should adapt their logic and recommendations according to the specific responsibilities of each pipeline stage. Rather than applying generic detection rules, tools should incorporate domain knowledge of ML practices to identify performance-critical smells and guide developers toward relevant optimizations. This ensures that the focus remains not only on code maintainability but also on runtime performance and resource efficiency.

\subsubsection*{Create Workflow-Integrated Assistants}
Our study points to a clear opportunity for the development of tools that integrate seamlessly into existing workflows while remaining aware of ML pipeline semantics. Developers would benefit from lightweight, IDE-integrated assistants capable of detecting performance smells, providing contextual explanations, and visualizing smell distribution across the pipeline. Such tools could offer stage-specific refactoring suggestions and help track regressions related to performance over time.

\subsection{Implications for Researchers}
\subsubsection*{Hybrid Classification for Pipeline Stage Identification}
To support our smell analysis, we introduced a hybrid classification method that combines zero-shot learning with domain-specific keyword matching. This approach enhances robustness by bridging semantic generalization with symbolic interpretability. While large pre-trained models like facebook/bart-large-mnli offer flexibility and cross-domain generalization, they may struggle with ambiguous or multi-purpose code files. The addition of ML-specific keywords helps refine predictions, particularly in cases where semantic signals are weak or conflicting. This hybrid design strikes a balance between expressiveness and precision, making it well-suited for multi-stage classification tasks. Researchers can adopt this strategy to reduce manual labeling efforts and improve the scalability of repository mining efforts.

\section{Threats to Validity} \label{sec:threats}
We now discuss threats to the validity of our study.

\textbf{Construct Validity}
A potential threat to the validity of our results stems from the uneven distribution of projects across application domains. This imbalance may introduce bias by amplifying the influence of overrepresented domains. To mitigate this risk, we applied several normalization strategies. First, performance smell occurrences were normalized per file and KLOC, ensuring that variations in codebase size across projects do not disproportionately impact the analysis. Second, all distributions were computed and reported as percentages rather than absolute counts, allowing for fairer comparisons across pipeline stages regardless of the domain prevalence. These normalization steps help reduce the structural imbalance in the dataset and support the robustness of our findings.

%\LEUSON{Use paragraphs. I've gotten lost many times while reading it. The text should be organized in a way that would be reviewed by one of the journal's reviewers.}
To evaluate performance smell frequency, we employed RIdiom due to its comprehensive smell detection capabilities and proven reliability \cite{zhang2022making}. We validated the tool's effectiveness through a two-stage manual process.
First, we examined a subset of 50 code snippets where RIdiom detected performance smells. Each detection was manually verified and classified as either a true positive (correct detection) or false positive (incorrect detection).
Second, to assess RIdiom's ability to correctly identify clean code, we manually reviewed 50 randomly selected files where RIdiom reported no smells. Files genuinely lacking smells were labeled as true negatives, while those containing overlooked smells were marked as false negatives.
This evaluation yielded the following metrics: RIdiom achieved 100\% precision%(all reported smells were accurate with no false positives)
, 77.78\% recall%(successfully detected most but not all actual smells)
, 85.86\% overall accuracy and an F1 score of 0.875. These results indicate that while RIdiom is highly precise, improvements in detection coverage would help reduce false negatives. %\LEUSON{pay attention how you report this. The previous reviewer left a comment about it.}

\textbf{Internal Validity}
Our classification approach combines zero-shot BERT-based classification with keyword heuristics, showing promising results despite certain limitations. This hybrid method enhances generalizability while capturing task-specific cues from code and context. However, challenges persist in scenarios requiring deeper semantic understanding or when ambiguous keywords introduce noise. While the model performs well overall, its 73\% precision for Model Deployment introduces a risk of false positives. To address this, our analysis prioritizes samples with high classification confidence. In this refined subset, \textit{Assign Multi Targets} consistently emerges as the dominant performance smell, while \textit{Truth Value Test} occurrences in broader predictions likely reflect noise from misclassified samples. We report these findings cautiously and recommend further refinement of the Model Deployment classifier.

An additional threat to internal validity comes from our keyword-based classification component. We initially built our keyword set using popular ML libraries and tutorials, then refined it based on our dataset's distribution and content. This iterative process improves precision but risks overfitting to our specific data. To mitigate this, we favored general-purpose terms (e.g., train, predict) and complemented keyword matching with structural code patterns, ensuring classifications rely on both terminology and code semantics.

A notable validity threat stems from assuming that performance smells in a file impact all associated pipeline stages. This is particularly problematic for files spanning multiple stages. Our current methodology aggregates smell occurrences at the file level and distributes them across all associated stages, potentially inflating or diluting certain smells' presence. For example, a problematic code in a Model Evaluation utility function might be incorrectly attributed to Data Processing if the file also contains preprocessing logic. This creates a misleading impression of uniform smell distribution. To mitigate this risk, our analysis for RQ$_3$ focuses specifically on mono-labeled files, ensuring performance smells are more tightly coupled with a single pipeline stage, reducing attribution noise and improving reliability of stage-specific distributions.
 
%\LEUSON{The next discussion is related to Table 9. When you presented it, you barely discussed such a table. So, bring the current discussion and place it there. Here, you just have to acknowledge the threats, possible aims to avoid them, and the impact on the results.}
\textbf{External Validity}
There are several potential threats to the generalizability of our results. First, our sample consists of open-source Python projects hosted on GitHub, which may not fully represent the entire range of ML projects. To enhance diversity, we employed various strategies to include different types of ML projects. Second, the performance smells assessed in this study are limited to those supported by the RIdiom tool, meaning that additional tools would be required to cover a broader range of smells. Furthermore, when classifying ML stages, we relied on external libraries that may not encompass all libraries commonly used in ML development. To mitigate this limitation, we focused on selecting the most widely used and relevant libraries for ML projects~\cite{nguyen2019machine}.

%\LEUSON{Why do you think that's an external threat? I think it's internal. You can use GPT when writing. Ask to evaluate your text, whether it captures the message you're trying to give.}

%\input{related}
\section{Related Work} \label{sec:related}
%\LEUSON{We have to provide a very detailed discussion of related studies. Maybe, relying on performance smells in general, bringing green SE into discussion, etc. But provide more content and details, compared with the findings reported here.}

Several studies have examined the relationship between code smells and performance, often highlighting how certain code smells can lead to inefficiencies in resource utilization \cite{zhang2022making}. However, no prior work investigated the distribution and comparison of these smells between ML and non-ML projects, as we do here. In this section, we review related work across three perspectives: (i) performance-related smells in general software systems, (ii) comparisons of code quality between ML and non-ML systems, and (iii) performance challenges across the stages of the ML pipeline.

\begin{comment}    
Several studies have examined the relationship between code smells and performance, often highlighting how certain code smells can lead to inefficiencies in resource utilization. 
However, overall, no prior work investigated the distribution and comparison of these smells between ML and non-ML projects, as we do here. 
For instance, Gesi \etal \cite{gesi2022codesmellsmachinelearning} identified code smells in machine learning (ML) systems, noting that poor architectural decisions and improper data management could lead to performance bottlenecks. 
Similarly, other researchers have looked at code smells like long methods or deeply nested conditionals, which can increase computational overhead~\cite{gesi2022codesmellsmachinelearning}. Still, their work stops short of defining and studying performance smells explicitly.
\end{comment}

\begin{comment}    
\subsection{Code Quality in ML vs. Non-ML Systems}
A growing body of work compared code quality between ML and non-ML systems. Hadhemi \etal analyzed code smells across both types of projects, identifying structural issues that could indirectly affect performance by making code more complex and difficult to maintain \cite{Hadhemi}. However, their work does not distinguish between the types of inefficiencies that arise %specifically in ML projects compared to non-ML ones, 
nor does it explicitly investigate performance smells as an isolated category. This gap leaves room to explore how different project types might uniquely influence the occurrence and severity of performance smells.
\end{comment}

\subsection{Performance Smells and Green Software Engineering}
Code smells have traditionally been associated with poor maintainability, but recent studies have highlighted their impact on runtime performance and energy consumption. For instance, Gesi et al.\cite{gesi2022codesmellsmachinelearning} identified code smells in machine learning (ML) systems, noting that poor architectural decisions and improper data management can lead to performance bottlenecks. Palomba et al.\cite{palomba2018diffuseness} further investigated how code smells affect code maintainability.

More recently, the concept of performance smells has gained attention. While Gesi et al.~\cite{gesi2022codesmellsmachinelearning} focused on code smells in ML systems, they did not specifically isolate performance smells as a distinct category. Our study extends this work by defining performance smells and comparing their prevalence in both ML and non-ML projects.

Simultaneously, the emerging field of Green Software Engineering (Green SE) emphasizes the importance of energy-efficient programming practices. Strubell et al.\cite{strubell2019energypolicyconsiderationsdeep} demonstrated that training large models like BERT on GPUs incurs high computational costs and environmental impacts. For example, training BERT on GPUs can generate a carbon footprint comparable to a transatlantic flight \cite{strubell2019energypolicyconsiderationsdeep}. They highlight the need for more efficient hyperparameter optimization techniques, such as Bayesian optimization, to minimize redundant computations during model development. Other studies have addressed these challenges by exploring more powerful hardware or distributed training techniques\cite{7723730, 10.1145/3377454}.

Performance smells pose a threat not only to speed and memory usage but also to sustainability. Previous studies have often overlooked the role of inefficiencies at the code level. By analyzing performance smells in Python ML projects, our study emphasizes how optimizing code quality can complement hardware and algorithmic improvements, offering new insights into enhancing both performance and energy efficiency.

\subsection{Code Quality in ML vs. Non-ML Systems}
A growing body of research compares code quality between ML and non-ML systems. Hadhemi et al.~\cite{Hadhemi} analyzed code smells in both contexts, identifying structural issues that could indirectly affect performance. However, their study does not differentiate the types of inefficiencies that arise in ML projects compared to non-ML ones, nor does it investigate performance smells as an explicit category.

Zhang et al.\cite{zhang2022making} addressed Pythonic idioms and introduced the RIdiom tool to automatically refactor non-idiomatic Python code. They showed that Pythonic constructs improve both readability and execution speed. Zid et al.~\cite{zid2024list} highlighted that many Python developers are unaware of efficient constructs such as generator expressions or list comprehensions, leading to subtle but impactful performance degradation. While such studies hint at performance-related issues, they do not examine their distribution across different project types, which is the focus of our work.

\subsection{Challenges Across ML Pipeline Stages}
Few studies have addressed performance issues across the distinct stages of the ML pipeline. Cao et al.~\cite{cao2022understanding} showed that bottlenecks often emerge in early pipeline stages such as data preprocessing and model training, due to resource-intensive operations. Our study complements these findings by mapping specific performance smells to the corresponding ML pipeline stages.

Unlike prior work, we provide a stage-wise breakdown of where performance smells are most frequent, revealing that Data Processing is particularly affected. Moreover, we observe that performance smells are not confined to training or preprocessing, but persist into deployment phases as well. This broader perspective helps stakeholders identify where optimizations are most needed across the ML development lifecycle.

\section{Conclusion and Future Work} \label{sec:conclusion}
This paper examines the prevalence of performance smells in both ML and non-ML Python projects. Our analysis of 300 GitHub projects reveals that ML projects are particularly susceptible to performance smells, especially those related to \textit{Assign Multi Targets}, \textit{Truth Value Test}, and \textit{Chain Compare}, which are often linked to the data-intensive and computationally complex nature of ML workflows. We also identify Data Processing as the stage the most impacted by these performance inefficiencies. The presence of performance smells in the Model Deployment stage shows that these smells are not restricted to the early stages of the ML pipeline. 

The insights from this research highlight the importance of tailoring performance optimization strategies to the unique demands of ML workflows. By addressing specific smells prevalent in ML projects, developers can enhance both the efficiency and maintainability of their code. Future work could focus on assessing the impact of these performance smells on the training time of models.
This research emphasizes the critical role of performance smells in ML project efficiency and provides a foundation for future work aimed at quantifying their impact on model training times. Future studies could focus on refining techniques for detecting and mitigating performance smells, with the goal of further improving Python-based ML systems' performance and maintainability.

\section{Declarations}
\label{sec:declaration}

\subsection{Funding} 
Not applicable.

\subsection{Ethical approval} 
Not applicable.

\subsection{Informed consent} 
Not applicable.

\subsection{Author Contributions}
François Belias designed the study, collected and processed the data, performed the calculations and the analysis of the results, and wrote the manuscript. Leuson Da Silva contributed to the study design, project classification and filtering, and provided critical feedback to improve the manuscript. Foutse Khomh supervised the project and provided overall guidance throughout the research. Cyrine Zid contributed to the filtering and classification of projects. All authors have read and approved the final version of the manuscript.

\subsection{Data Availability Statement} 
All the data used for this study can be found in our replication package \cite{MLvsNO_ML_replication}

\subsection{Conflict of Interest}
The authors declare no conflict of interest.

\subsection{Clinical trial number} 
Not applicable.

\bibliographystyle{plain}
\bibliography{main}
%\usepackage[style=numeric]{biblatex}
%\addbibresource{main}
%\printbibliography

\end{document}